\documentclass{emulateapj}
\usepackage{natbib}
\usepackage{apjfonts}
\usepackage{epsfig}

\newcommand{\etal}{et al.}

\def\km{{\rm\thinspace km}}
\def\kpc{{\rm\thinspace kpc}}

\def\Mpc{{\rm\thinspace Mpc}}
\def\Msun{\hbox{$\rm\thinspace M_{\odot}$}}

\def\s{{\rm\thinspace s}}
\def\yr{{\rm\thinspace yr}}

\def\ltsima{$\; \buildrel < \over \sim \;$}
\def\simlt{\lower.5ex\hbox{\ltsima}}
\def\gtsima{$\; \buildrel > \over \sim \;$}
\def\simgt{\lower.5ex\hbox{\gtsima}}

\def\mdot{\hbox{$\dot m$}}
\def\<{\thinspace}
\def\ss{\s\ }           
\def\Mpc{{\rm\thinspace Mpc}}
\def\km{{\rm\thinspace km}}

\def\s{{\rm\thinspace s}}

\def\kmps{\hbox{$\km\s^{-1}\,$}}
\def\kmpspMpc{\hbox{$\kmps\Mpc^{-1}$}}

\newcommand\beq{\begin{equation}}
\newcommand\eeq{\end{equation}}
\newcommand\beqa{\begin{eqnarray}}
\newcommand\eeqa{\end{eqnarray}}

\shorttitle{Direct cosmological simulations of the growth of black
holes and galaxies} \shortauthors{Di Matteo \etal}
\slugcomment{Submitted to ApJ 05/14/07}
\begin{document}

\title{Direct cosmological simulations of the growth of black holes
and galaxies} \author{ Tiziana Di Matteo,\altaffilmark{1} J\"org
Colberg,\altaffilmark{1} Volker Springel,\altaffilmark{2} Lars
Hernquist\altaffilmark{3} \&\ Debora Sijacki\altaffilmark{2} }
\altaffiltext{1} {Physics Department, Carnegie Mellon University, 5000
Forbes Avenue, Pittsburgh, PA 15213}
\altaffiltext{2}{Max-Planck-Institut f\"{u}r Astrophysik,
Karl-Schwarzchild-Stra\ss e 1, 85740 Garching bei M\"{u}nchen,
Germany} \altaffiltext{3}{Harvard-Smithsonian Center for Astrophysics,
60 Garden Street, Cambridge, MA 02138, USA}

\begin{abstract}
  We investigate the coupled formation and evolution of galaxies and their
  embedded supermassive black holes using state-of-the-art hydrodynamic
  simulations of cosmological structure formation. For the first time, we
  self-consistently follow the dark matter dynamics, radiative gas cooling,
  star formation, as well as black hole growth and associated energy feedback
  processes, starting directly from initial conditions appropriate for the
  $\Lambda$CDM cosmology.  Our modeling of the black hole physics is based on
  an approach we have recently developed and tested in simulations of isolated
  galaxy mergers. Here we apply the same model in cosmological simulations to
  examine: (i) the predicted global history of black hole mass assembly in
  galaxies, (ii) the evolution of the local black hole-host mass correlations
  and (iii) the conditions that allow rapid growth of the first quasars,
  as well
  as the properties of their hosts and descendants today. We find that our
  simulations produce a total black hole mass density $\rho_{\rm BH}\simeq 2
  \times 10^{5} \Msun\Mpc^{-3}$ by $z=0$, 
  in good agreement with observational
  estimates. The black hole accretion rate density, $\dot{\rho}_{\rm BH}$,
  peaks at lower redshift and evolves more strongly at high redshift than the
  star formation rate density, $\dot{\rho}_{*}$, with an approximate scaling
  as $\dot{\rho}_{\rm BH}/\dot{\rho}_{*} \propto (1+z)^{-4}$ at $z\ge3$. On
  the other hand, the ratio $\rho_{\rm BH}/\rho_{*} \sim (1+z)^{-0.6}$ of
  black hole to stellar mass densities shows only a moderate evolution at low
  redshifts $z \lesssim 3$. For the population of galaxies identified in the
  simulations at $z=1$ we find strong correlations between black hole mass and
  velocity dispersion or mass of the stellar systems. The predicted
  correlations agree well with the measured local $M_{\rm BH}-\sigma$ and
  $M_{\rm BH} -M_{*}$ relationships, but also suggest a weak evolution with
  redshift in the normalization, and in particular the slope. However, the
  magnitude of this effect is sensitive to the range of masses being probed.
  For
  stellar masses of $M_{*}\ge 3\times 10^{10}$, we predict a trend of
  increasing $M_{*}/ M_{\rm BH}$ with redshift, in agreement with recent
  direct estimates of the BH to host stellar mass ratio at high redshift and
  the conjecture that a more fundamental relation (a BH fundamental plane)
  should involve both $M_{*}$ and $\sigma$.  We find that our simulation
  models can also produce quite massive black holes at high redshift, as a
  result of extended periods of exponential growth in relatively isolated,
  rare regions that collapse early and exhibit strong gas
  inflows. Interestingly, when followed to their descendants, these first
  supermassive BH systems are not necessarily the most massive ones today,
  since they are often overtaken in growth by quasars that form later.
\end{abstract}


\section{Introduction\label{sec:intro}}
Following the discovery of quasars \citep{Schmidt1963, Greenstein1963} it was
suggested that supermassive black holes ($10^6-10^9$~\Msun) lie at the centers
of galaxies, and that the quasar activity is fueled by the release of
gravitational energy from their accreted matter. The remnants of quasar phases
at early times are probably the supermassive black holes found at the centers
of galaxies in our local Universe. Interestingly, the properties of these
supermassive black holes are tightly coupled to the mass \citep{Magorrian1998}
and velocity dispersion of their host galaxies, as manifested in the $M_{\rm
  BH}-\sigma$ relation of spheroids \citep{Ferrarese2000, Gebhardt2000}. In
addition, the black hole mass is correlated with the concentration or Sersic
index \citep{Graham2006}.  Most recently, \citet{Hopkins2007a} have shown 
that these various correlations are not independent, and can be
understood as projections of a ``black hole fundamental plane'' (BHFP),
similar to that describing properties of elliptical galaxies.

The existence of highly luminous quasars also
constrains the formation and evolution of massive galaxies and the epoch of
reionization. Quasars with inferred black hole masses in excess of
$10^9$~\Msun\ have now been discovered out to $z\sim 6$ \citep{Fan2003},
indicating an early formation time for black holes and galaxy spheroids and
posing a significant challenge for theoretical models of high-redshift quasar
and galaxy formation.

This growing observational evidence, drawn from local galaxies to high
redshift quasars, argues for a close connection between the formation and
evolution of galaxies and of their central supermassive black holes. However,
the physical nature of this relationship has yet to be understood in detail.
Indeed, there are significant gaps in our observational and theoretical
knowledge of the history of black hole formation and evolution in galaxies.

For example, current velocity dispersion measurements are inconclusive about
the important question whether the tight scaling relations evolve with
redshift \citep{Woo2006,Shields2006}, or are essentially invariant
\citep{Shields2003} as a function of time. We note that some
evolution in the ratio of black hole to halo mass
is suggested by clustering constraints \citep[e.g][]{Adelberger2005,
Lidz2006}, but these measurements do not directly address
the relationship between black hole and properties of the
luminous host galaxy.  More relevant are 
comparisons of the black hole mass inferred from quasar observations
to the host stellar mass, both observationally
\citep{Merloni2004} and theoretically \citep{Hopkins2006a},
which indicate an evolution in e.g. the Magorrian relation, in
the sense that black holes are more massive relative to luminous
spheroids at high redshifts than at $z=0$.  

Theoretical studies of the co-evolution of black holes and galaxies have so
far mostly used so-called semi-analytical modeling
\citep[e.g.][]{Kauffmann2000, Cattaneo1999, Wyithe2003a, Volonteri2003,
DiMatteo2003, Granato2004, Springel2005c, Cattaneo2005, Croton2006,
DeLucia2006, Malbon2007} of galaxy formation, in which the growth of
galaxies and their embedded black holes is followed with simple physical
parameterizations on top of dark matter merging history trees. Many of these
models assume that quasar activity is triggered by major galaxy mergers,
motivated by hydrodynamical simulations that have shown that gravitational
tidal fields during major mergers of gas rich galaxies produce strong gas
inflows \citep{BarnesHernquist1991, BarnesHernquist1996}, which lead to a
burst of nuclear star formation \citep{Mihos1996} and are likely the
prerequisite for rapid black hole growth and quasar activity.  Nearby quasars
are indeed preferentially found in tidally disturbed objects
\citep[e.g][]{Jogee2004}, corroborating the importance of galaxy interactions
and mergers for major black hole growth.

Many theoretical explanations for the observed correlations between galaxy
properties and black hole mass rely on some form of self-regulated growth of
the BHs. For example, it has been suggested that the central black holes grow
until they release sufficient energy to unbind the gas that feeds them from
their host galaxy \citep{CiottiOstriker1997, Silk1998, Fabian1999,
Wyithe2003a}. We recently explored such a local energy feedback for the first
time with self-consistent, fully three-dimensional hydrodynamic simulations of
galaxies \citep{DiMatteo2005,Springel2005a} that include a treatment of
accretion on supermassive black holes and their associated energy
feedback. These simulations have demonstrated that the fundamental BH-host
correlation including the $M_{\rm BH}-\sigma$ relation can indeed be
reproduced in feedback-regulated models of BH growth \citep{DiMatteo2005}, in
accordance with theoretical conjectures. At the same time, the dynamical
coupling in the simulations of hydrodynamical gas inflow, star formation,
black hole growth and associated feedback processes gives them substantial
predictive power well beyond that of simplified analytical and semi-analytical
models. Besides the $M_{\rm BH}-\sigma$ or $M_{\rm BH}-M_{*}$
\citep{DiMatteo2005, Robertson2006a} relationships, the simulation models can
for example predict the detailed properties of the spheroidal galaxies forming
in major mergers and how they correlate with the BH masses. In fact, they
suggest the existence of a fundamental plane relation for BHs ($M_{BH} \propto
\sigma^{3.0} R_e^{0.5}$ \citep{Hopkins2007a}, provide an explanation for the
red colors of massive elliptical galaxies \citep{Springel2005b}, and
describe the fundamental plane of elliptical galaxies \citep{Robertson2006b}.
They also suggest luminosity-dependent quasar lifetimes, leading to a new
interpretation for the origin of the quasar luminosity function and its
evolution over cosmic history~\citep{Hopkins2005, Hopkins2006a}.

In the present paper, we extend these earlier studies by carrying out fully
cosmological hydrodynamic simulations of the $\Lambda$CDM model that jointly
follow the growth of galaxies and supermassive black holes, as well as their
associated feedback processes. Our approach is based on the same methodology
that we have developed and applied in the high-resolution simulations of
galaxy mergers, augmented with a suitable mechanism to seed emerging new dark
matter halos with a small black hole that can then grow by gas accretion later
on. While much more restricted in numerical resolution than simulations of
individual galaxy mergers, our modeling of star formation and black hole
physics in terms of a sub-resolution treatment provides quite accurate results
already at comparatively coarse resolution, an important prerequisite for
attempting to model these processes in cosmological simulations. Nevertheless,
numerical resolution is clearly an important limitation of our cosmological
results, an aspect that we will discuss in more detail where appropriate. With
this caveat in mind, we would like to stress however that the unambiguous
initial conditions of direct cosmological simulation make them in principle
the most powerful and accurate tool for studying the interplay of galaxy
formation and black hole growth.  Our aim is therefore to examine how well our
current model for treating BH physics in simulations does in present
state-of-the-art hydrodynamical calculations of cosmic structure formation,
and what we can learn from them to advance our theoretical understanding of
the co-evolution of galaxies and supermassive black holes. In this study we
shall focus on basic properties of the black hole population, like the
evolution of the cosmic BH mass density, and the correlations between BH
masses and host galaxy properties. In \citet{Sijacki2007} we also study an
extension of our feedback model with a `radio mode' that is active at low
accretion rates and is distinct from the normal quasar activity, allowing us
to study the formation of AGN in rich galaxy clusters at low redshifts, where
it is likely to be important.

This paper is structured as follows.  In \S2 we describe our simulation set
and the numerical modeling adopted for the ISM, star formation and gas
accretion onto black holes. In \S3 we present our results for the evolution of
the global black hole mass density and compare it to the cosmic history of
star formation and the evolution of the stellar density.  In \S4 we examine our
results for the fundamental BH/host correlations measured from the simulation
and for their evolution from high to low redshift. Finally, we summarize and
discuss our findings in \S5.

\section{Methodology\label{sec:method}}

\subsection{Numerical code}
In this study we focus on a $\Lambda$CDM cosmological model with parameters
chosen according to the first year results from the Wilkinson Microwave
Anisotropy Probe \citep[WMAP1;][] {Spergel2003}, $\Omega_0 = 0.3$,
$\Omega_{\Lambda} = 0.7$, Hubble constant $H_0 = 100 h \kmpspMpc$ with $h
=0.7$ and a scale invariant primordial power spectrum with index $n=1$, with a
normalization of the amplitude of fluctuations $\sigma_{8} = 0.9$.
\footnote{The largest simulation presented here had already been started by
  the time the updated third year constraints have become available
  \citep[WMAP3;][]{Spergel2006}. We comment on effects on the growth of the
  halo mass function owing to the lower amplitude of fluctuations, $\sigma_{8}$
  implied by WMAP3 in ~\citet{Li2007, Sijacki2007}}. We use a significantly
  extended version of the parallel cosmological TreePM-SPH code {\small
  GADGET2}~\citep{Springel2005d} to evolve a realization of $\Lambda$CDM
  initial conditions from high to low redshift. The combination of a
  high-resolution gravitational solver with individual and adaptive timesteps
  allows this code to bridge a large dynamic range both in length- and
  timescales.  Gas dynamics is followed with the Lagrangian smoothed particle
  hydrodynamics (SPH)~\citep[e.g] []{Monaghan1992} technique, which we employ
  in a formulation that manifestly conserves energy and entropy, despite the
  use of fully adaptive SPH smoothing lengths~\citep{Springel2002}.  Radiative
  cooling and heating processes are computed as in \citet{Katz1996}, with a
  spatially uniform photoionizing UV background that is imposed externally.

Within cosmological (or galaxy-sized) numerical simulations, it is presently
(and for some time to come) not feasible to follow the physics of star
formation and black hole accretion from first principles down to scales of
individual stars or black holes.  Any numerical model of galaxy formation
therefore needs to make substantial approximations for some of the relevant
physics on unresolved scales.

For modeling star formation and its associated supernova feedback we use the
sub-resolution multiphase model for the interstellar medium developed by
\citet{Springel2003a}. In this model, a thermal instability is assumed to
operate above a critical density threshold $\rho_{\rm th}$, producing a two
phase medium consisting of cold clouds embedded in a tenuous gas at pressure
equilibrium. Stars form from the cold clouds, and short-lived stars supply an
energy of $10^{51}\,{\rm ergs}$ to the surrounding gas as supernovae. This
energy heats the diffuse phase of the ISM and evaporates cold clouds, thereby
establishing a self-regulation cycle for star formation.  $\rho_{\rm th}$ is
determined self-consistently in the model by requiring that the equation of
state (EOS) is continuous at the onset of star formation. The cloud
evaporation process and the cooling function of the gas then determine the
temperatures and the mass fractions of the two hot and cold phases of the
ISM, such that the EOS of the model can be directly computed as a function of
density. The latter is encapsulating the self-regulated nature of star
formation owing to supernovae feedback in a simple model for a multiphase ISM.
As in the \citet{Springel2003a} model we have included a model
for supernova-driven galactic winds with an initial wind speed of $v \sim 480
\kmps$.

For the parameter settings adopted here, the model reproduces the observed
star formation rate surface densities in isolated spiral
galaxies~\citep{Kennicutt1989, Kennicutt1998}. Using a large number of nested
cosmological simulations, the approach we adopt (and the parameters we use)
has also been shown to lead to a numerically converged estimate
for the cosmic star formation history of the universe that agrees reasonably
well with low redshift observations \citep{Springel2003b,Hernquist2003}.
For the modeling
of BH accretion and feedback we adopt a similar strategy as for star
formation, which we discuss next.

\subsection{Accretion and Feedback from Supermassive black holes}

\begin{figure*}
\begin{center}
\resizebox{7.6cm}{!}{\includegraphics{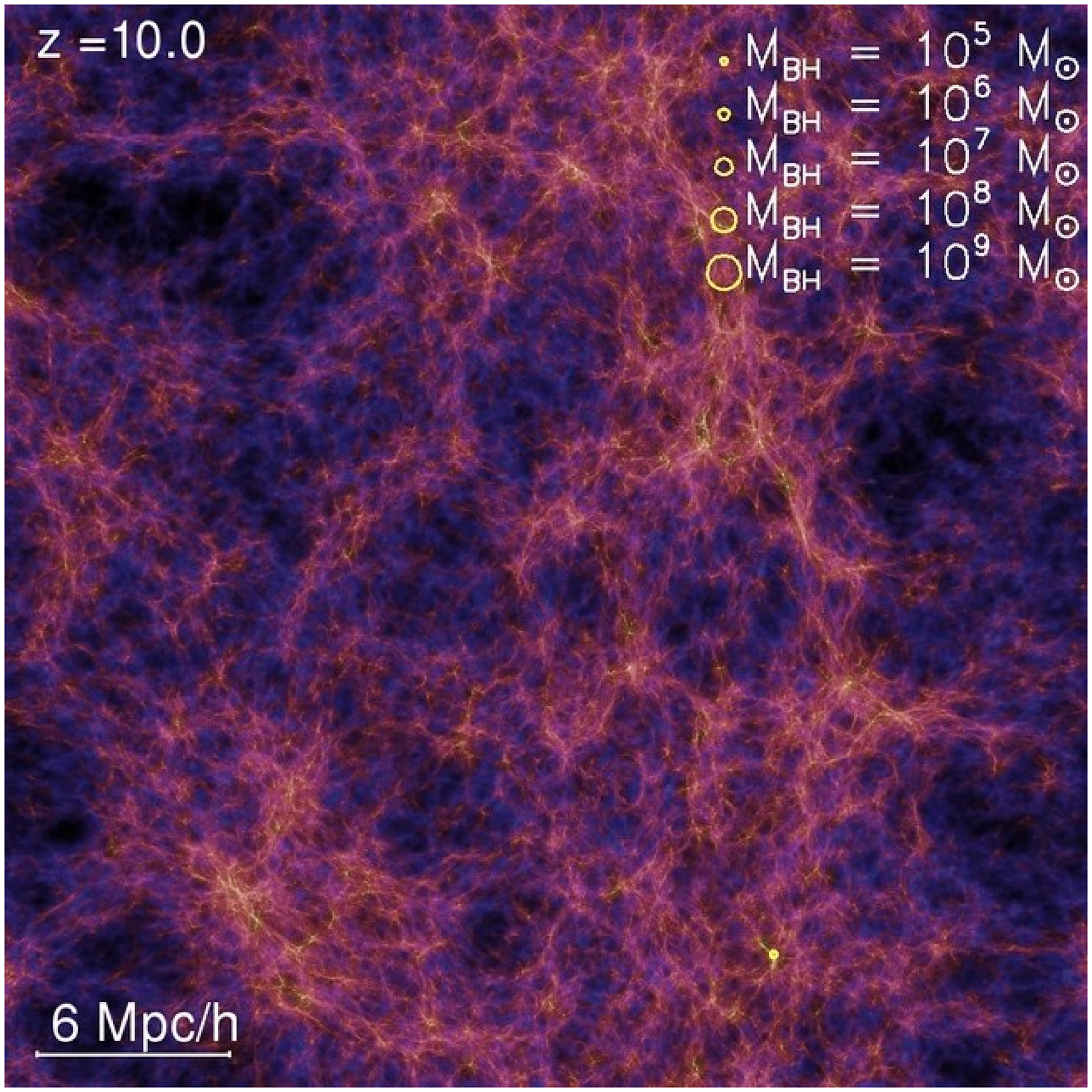}} %
\resizebox{7.6cm}{!}{\includegraphics{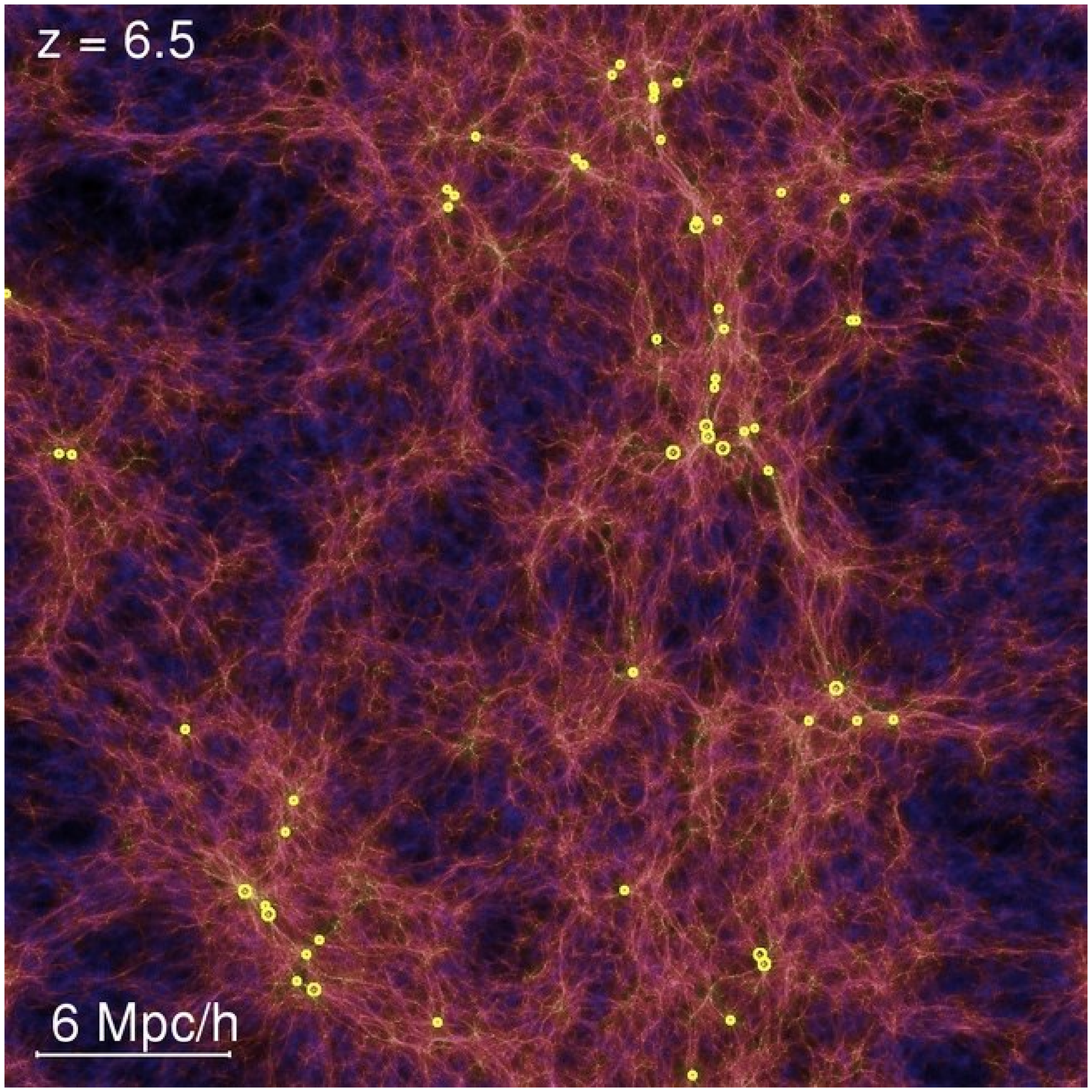}}\\
\resizebox{7.6cm}{!}{\includegraphics{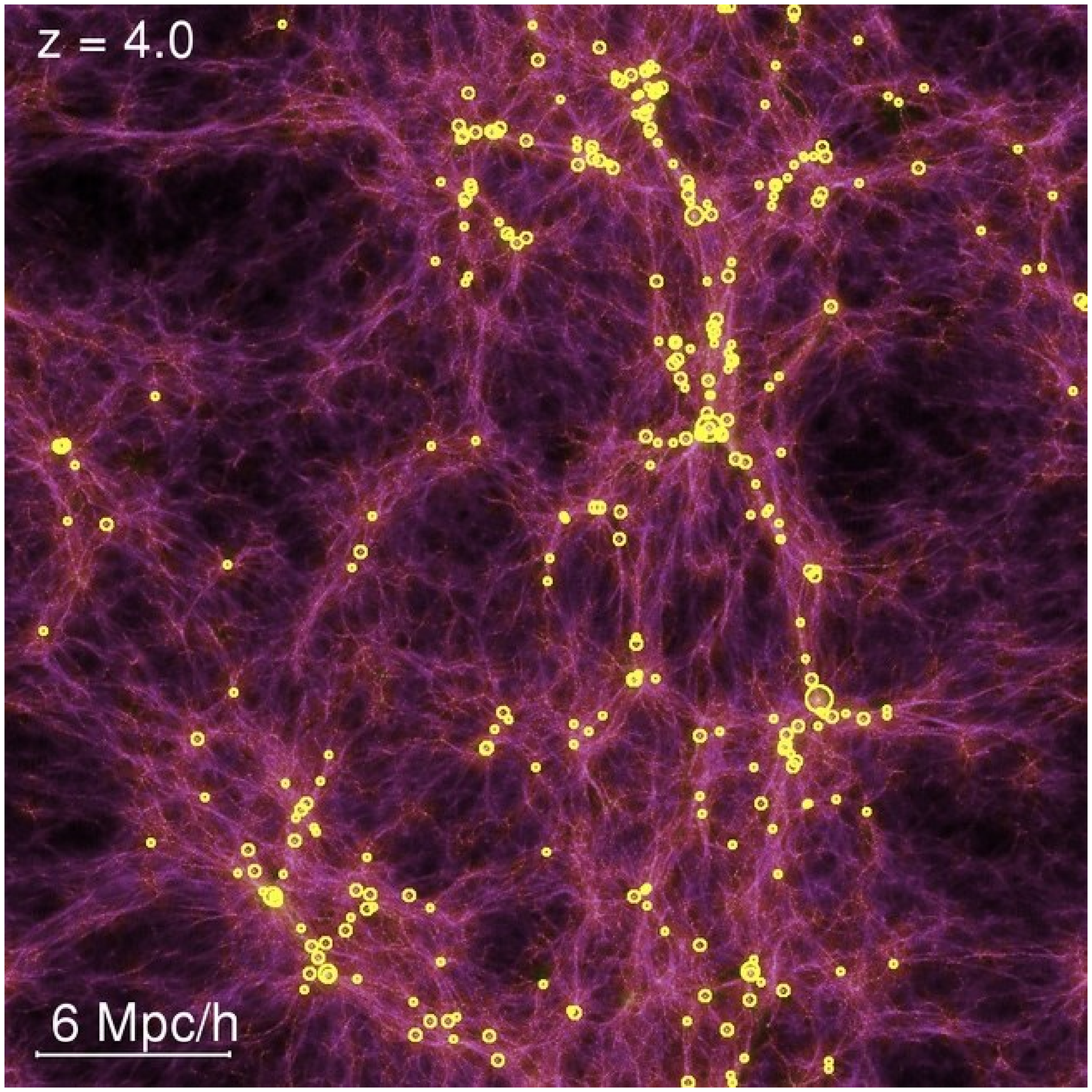}} %
\resizebox{7.6cm}{!}{\includegraphics{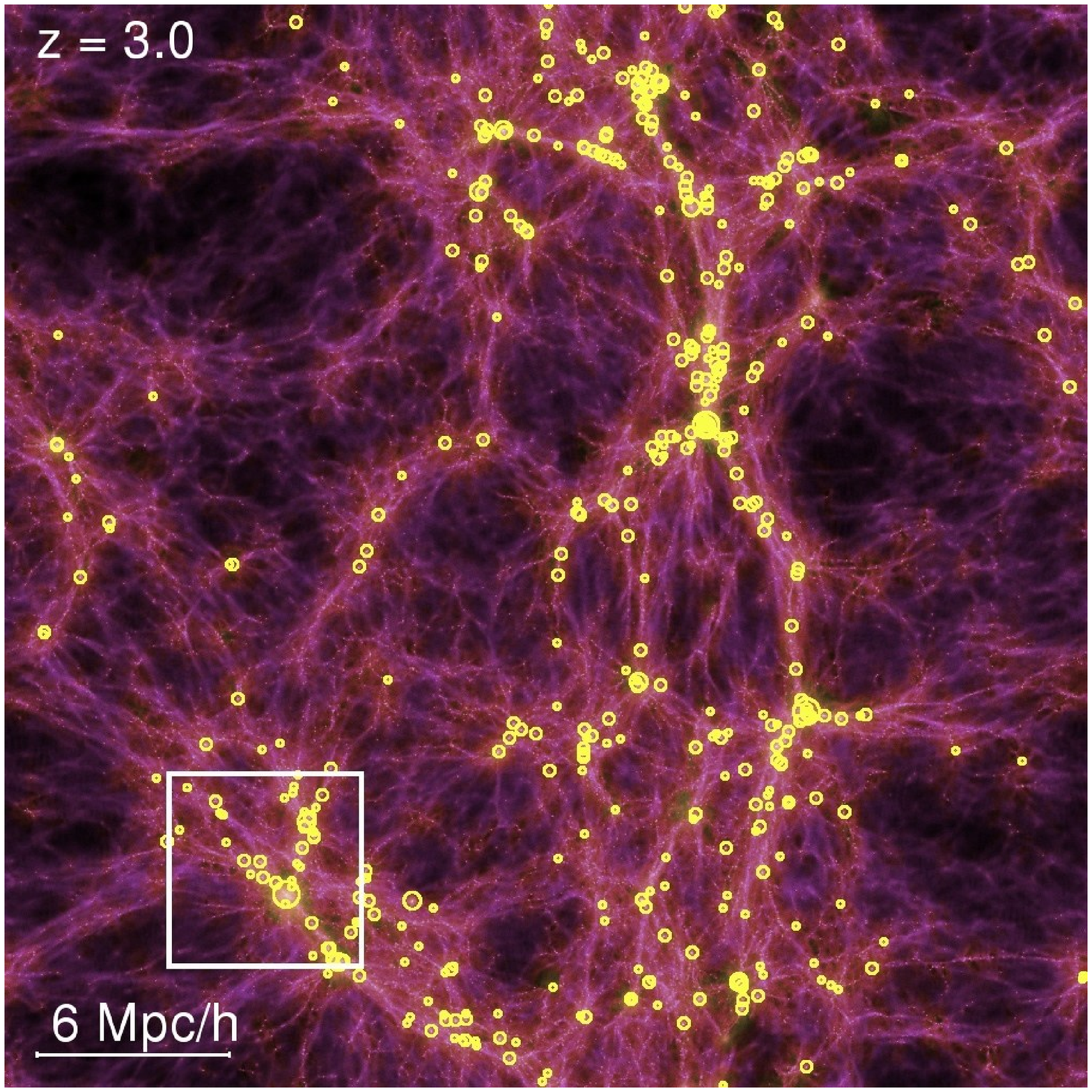}}\\
\resizebox{7.6cm}{!}{\includegraphics{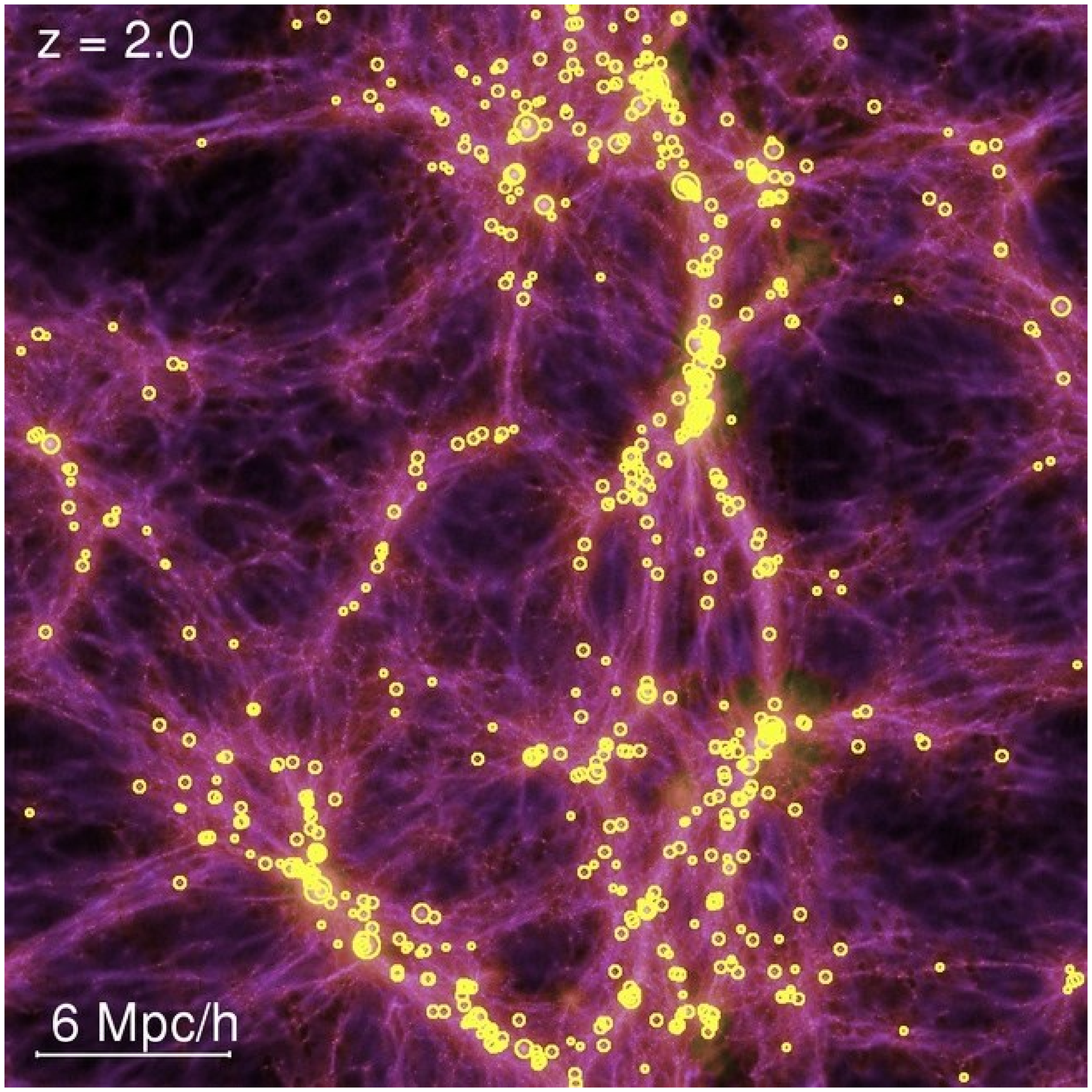}} %
\resizebox{7.6cm}{!}{\includegraphics{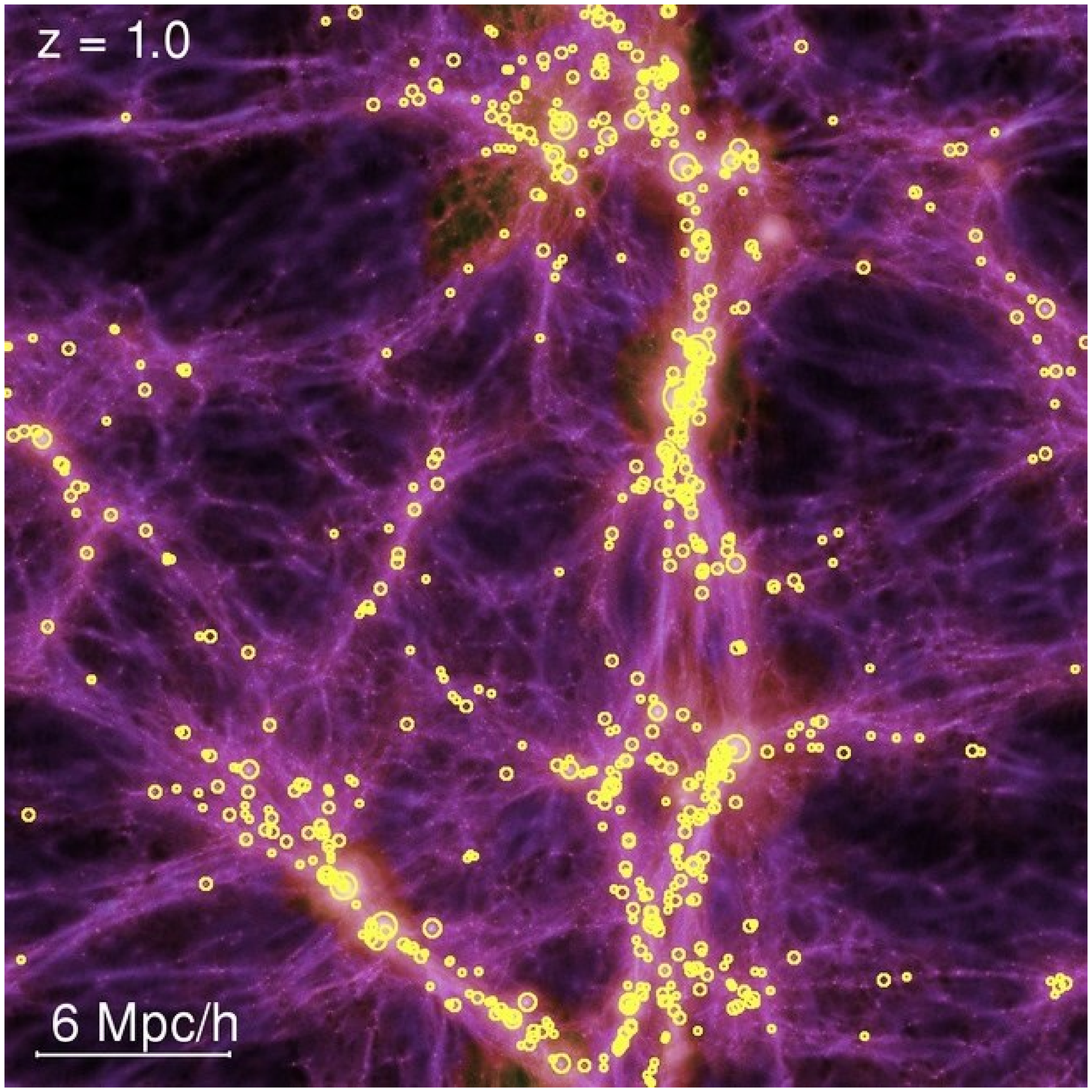}}
\end{center}
\caption{Projected baryonic density field in slices of thickness
$5000\,h^{-1}\kpc$ through our high resolution simulation, color-coded
by temperature and with brightness proportional to the logarithm of
the gas density. Each panel shows the same region of space at
different redshifts, as labeled.  The circles mark the positions of
the black holes, with a size that encodes the BH mass, as indicated in
the top left panel.}
\label{fig:D6slices}
\end{figure*}

As for star formation, we adopt a sub-resolution model to capture the main
features of accretion and associated feedback on supermassive black holes
\citep[as introduced in][]{Springel2005a, DiMatteo2005}.  To this end, we
represent black holes by collisionless `sink' particles that can grow in mass
by accreting gas from their immediate environments, or by merging with other
black holes. We estimate the gas accretion rate onto a black hole using a
Bondi-Hoyle-Lyttleton parameterization \citep{Bondi1952, BondiHoyle1944,
Hoyle1939}. In this description, the accretion rate onto the black hole is
given by $\dot{M}_{\rm B} \, = \, {{4\pi \, \, [G^2 M_{\rm BH}^2 \, \rho}] /
{(c_s^2 + v^2)^{3/2}}} \, $ where $\rho$, $c_s$ are the density and sound
speed of the ISM gas, respectively, and $v$ is the velocity of the black hole
relative to the surrounding gas. In this study we also allow the black hole to
accrete at mildly super-Eddington values and impose a maximum allowed
accretion rate equal to $2x$ Eddington rate, $\dot{M}_{\rm Edd}$. As we shall
see such conditions are generally achieved at high redshift.  We note that the
detailed relativistic accretion flow onto the black hole is {\em unresolved}
in our simulations, but if the limitating factor for rapid growth of BHs lies
in the larger-scale gas distribution around the black hole, {\em which is
resolved}, then the Bondi prescription should capture the dependence of the
mean accretion rate onto the conditions of the gas in the region around the
black hole.

The radiated luminosity, $L_{\rm r}$ is related to the gas accretion rate,
$\dot {M}_{\rm BH}$, by the radiative efficiency $ \eta \, = \, {{L_{\rm
r}}/({\dot {M}_{\rm BH} \, c^2}}) \,$, which simply gives the mass to energy
conversion efficiency, set by the amount of energy that can be extracted from
the innermost stable orbit of an accretion disk around a black hole. We will
adopt a fixed value of $\epsilon_{\rm r} =0.1$, which is the mean value for a
radiatively efficient \citep{Shakura1973} accretion disk onto a Schwarzschild
black hole.

We assume that some small fraction $\epsilon_{\rm f}$ of the radiated
luminosity $L_{\rm r}$ can couple to the surrounding gas in the form of
feedback energy, viz. $ \dot{E}_{\rm feed} \, = \, \epsilon_{\rm f} \, L_{\rm
  r}$
In accordance with our previous studies of galaxy merger simulations we take
$\epsilon_{\rm f} = 5 \% $.  This value governs the normalization of the
$M_{\rm BH} - \sigma$ relation, and $\epsilon_{\rm f} = 5 \% $ brings it into
agreement with current observations \citep{DiMatteo2005}. We note that
$\epsilon_{\rm f}$ is effectively the only free parameter in our black hole
model. After fixing it to reproduce the normalization of the observed $M_{\rm
  BH} - \sigma$ from the galaxy models we do not vary it in any of our
cosmological simulations.

For simplicity, we deposit the feedback energy isotropically in the region
around the black hole.  Lack of spatial resolution precludes us from
considering mechanical modes of releasing the energy, e.g.~in the form of a
jet. However, we note that is is plausible that such other forms of energy are
thermalized eventually as well, and that the final impact of the feedback
depends primarily on the total amount of energy released and less on the form
it is released in.  This is likely to be true generally, provided that
the energy (or momentum) is imparted to the surrounding gas on length
scales small and time scales short compared with those that characterize
the host galaxy.  In that event, the impact of black hole feedback will
be explosive in nature and, indeed, the blowout phase of evolution in
our simulations is well-described by a generalized Sedov-Taylor
blast-wave solution \citep{Hopkins2006b,HH2006}.
In any case, we emphasize that despite an isotropic release of the energy,
the response of the gas can still be decidedly anisotropic, e.g.~when a dense
gas disk is present that channels the gas response into a collimated outflow.

The idea that we follow with our feedback modeling here is the rapid
accretion phases of BHs at times close to their critical growth phases. Such
`quasar' phases are typically relatively short-lived and require galaxy
mergers to produce the strong gravitational tidal forcing necessary for
sufficient nuclear gas inflow rates. It is presently unclear whether the
accretion disks in such modes actually produce mechanical jets of release
their feedback in another way. It is however evident that at low redshift
(e.g.; $z < 1$ which we are not studying directly here) that when BHs that are
embedded in the hot gas atmospheres of large groups and clusters channel their
energy in mechanical feedback in the form of relativistic jets, generating
buoyantly rising radio bubble. In \citet{Sijacki2007} we consider this form
of feedback mode.


An important remaining question in our model concerns the ultimate origin of
the BHs. Since our accretion rate estimate can only grow a BH that already
exists, we assume that a physical process that produces small seed BHs is
operating sufficiently efficiently that effectively all halos above a certain
threshold mass contain at least one such seed BH. Whether or not they can then
grow to larger masses by gas accretion will be determined by the local gas
conditions, as described in our model above. In order to achieve such a
seeding at a technical level in cosmological runs we use an on-the-fly
`friends-of-friends' group finder algorithm which is called at intervals
equally spaced in the logarithm of the scale-factor $a$, with $\Delta \log{a} =
\log{1.25}$. This provides the locations and mass of all halos in the
simulation. If a halo is more massive than our threshold and does not contain
any black hole yet, we endow it with one by converting its densest gas
particle into a sink particle with a seed black hole mass of $10^5\, h^{-1}
\Msun$.  The further growth of the black hole sink can then proceed by gas
accretion, at a rate that depends sensitively on the local conditions, or by
mergers with other black hole sink particles.  The total cumulative black hole
mass introduced in this way as seeds is negligible compared to the mass growth
by gas accretion. We note that being able to run a fast, parallel
`friends-of-friends' algorithm on the fly during simulations is an important
technical prerequisite of our technique.

Further motivation for this choice of seeding procedure is based on the
currently proposed scenarios for the seed black holes in galaxies. To grow a
supermassive black hole to a mass of $\sim 10^{9} \Msun$ in less than a
billion years, as required by presence of the $z=6$ SDSS quasars, may require
(1) the catastrophic collapse of a supermassive star that forms a large
initial black hole of mass $10^4-10^6\Msun$ \citep{Carr1984, Bromm2003,
Begelman2006}, or (2) alternatively, smaller black hole seeds ($M \sim 10^2
\Msun$) may form from the first PopIII stars at $z \sim 30$ and grow
exponentially from then on \citep{Abel2002, Bromm2004, Yoshida2006}.
In our simulations,
black hole seeds of mass $M=10^5 \,h^{-1}\Msun$ are introduced into galaxies
as they initially reach $M_{\rm halo} = 10^{10}\,h^{-1} \Msun$.  This choice
is a good approximation to what is expected for both of the hypotheses
outlined above.  For (1) this is roughly in the correct range; whereas for (2)
Eddington growth predicts that the black hole has grown to roughly these
values by the time of collapse of $M \sim 10^{10}h^{-1} \Msun$ perturbations,
which occurs at $z \sim 10$ in our standard cold dark matter scenario.
Additionally, although not required, this value of the initial black hole mass
to galaxy ratio fits the observed relations at low redshift
\citep{Magorrian1998, Ferrarese2000}. It is important to note that the
dominant growth of black holes always occurs in exponential Eddington phases
induced by gas accretion so that our results are rather insensitive to the
specific choice of the seed mass.

\begin{table}
\begin{center}
\caption{Numerical parameters of cosmological simulations with BHs}
\label{tab:simul}
\begin{tabular}{ccccccc}
\hline\hline
Run  &  Boxsize & $N_{p}$ & $m_{\rm DM}$ & $m_{\rm gas}$ & $\epsilon$& $z_{\rm en
d}$ \\
     &  $h^{-1}$Mpc &&  $h^{-1} \Msun$ &$h^{-1} \Msun$ &  $h^{-1}$ kpc &  \\
\hline
  D4 & 33.75 & $2\times 216^3$& $2.75 \times 10^{8}$ & $4.24 \times 10^{7}$  & 6.25
& 0.00 \\
D6 ({\it BHCosmo}) & 33.75 & $2\times 486^3$& $2.75\times 10^{7}$ &  $4.24\times 10^{6}$&
2.73 & 1.00 \\
E6 & 50 & $2\times 486^3$& $7.85\times 10^{7}$ &  $1.21\times 10^{7}$&
4.12 & 4.00 \\
\hline\\
\end{tabular}
\end{center}
\vspace{-1cm}
\end{table}

When galaxies and their surrounding dark halos merge to form a single dark
matter halo, their central black holes are also expected to merge eventually,
so hierarchical black hole mergers contribute to the growth of the central
black holes.  Whether the forming black hole binaries can really coalesce
efficiently is however a matter of debate.  In a stellar environment, it has
been argued that the binary hardens only very slowly \citep{Begelman1980,
Milosav2003}, while in gaseous environments binaries may coalesce rapidly
owing to strong dynamical friction with the gas \citep{Makino2004,
Escala2004}. In our galaxy-sized simulations, and even more so in the
cosmological boxes, it is not possible to treat in detail the problem of
binary hardening, nor to directly calculate the ejection of black holes by
gravitational recoil, or by three-body sling-shot ejection of black holes in
triple systems.  Because galaxies have typically large central concentrations
of gas we instead assume that two black hole particles merge quickly if they
come within the spatial resolution of the simulation and their relative speed
lies below the gas sound speed. In practice, this means that two sink
particles that fulfill these conditions are merged into a single BH particle,
with their masses combined.

\begin{figure*}
\begin{center}
\resizebox{16.0cm}{!}{\includegraphics{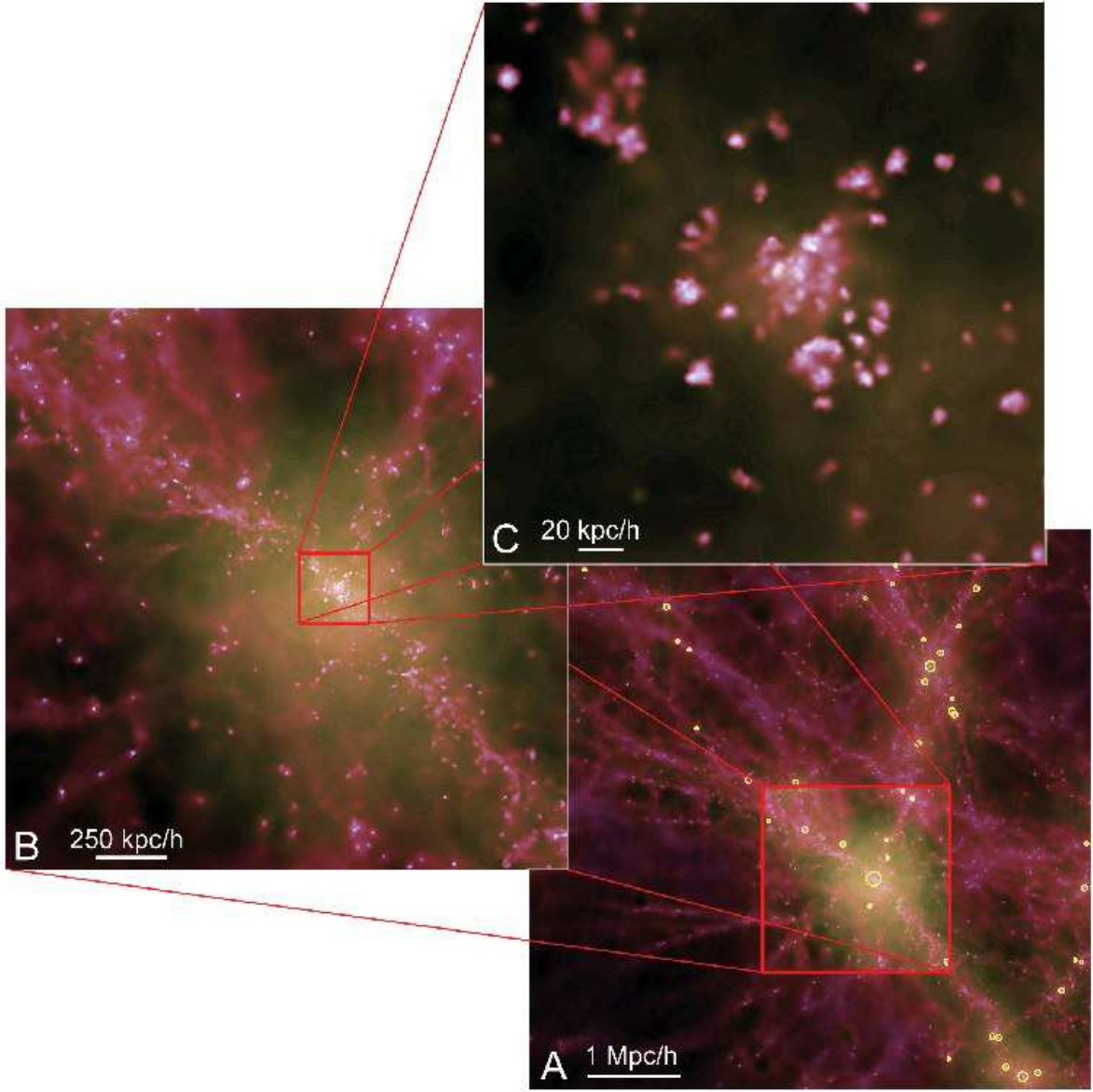}}
\end{center}
\caption{Three level zooms into the simulation region marked by the white
rectangle in the $z=3$ panel of Fig~\ref{fig:D6slices}.  The three panels show
the gas surface density, color-coded by temperature.  The panels show slices
of thickness 5000$ h^{-1} \kpc$ and of decreasing width (from A to C) as we
zoom into the region around a black hole of mass, $M_{BH} \sim 7\times
10^{7}$. The yellow circles in the bottom right panel show the black holes in
the region, with a symbol size that is related to the BH mass as in
Fig.~\ref{fig:D6slices}.}
\label{fig:D6zoomsgas}
\end{figure*}

\begin{figure*}
\begin{center}
\resizebox{16.0cm}{!}{\includegraphics{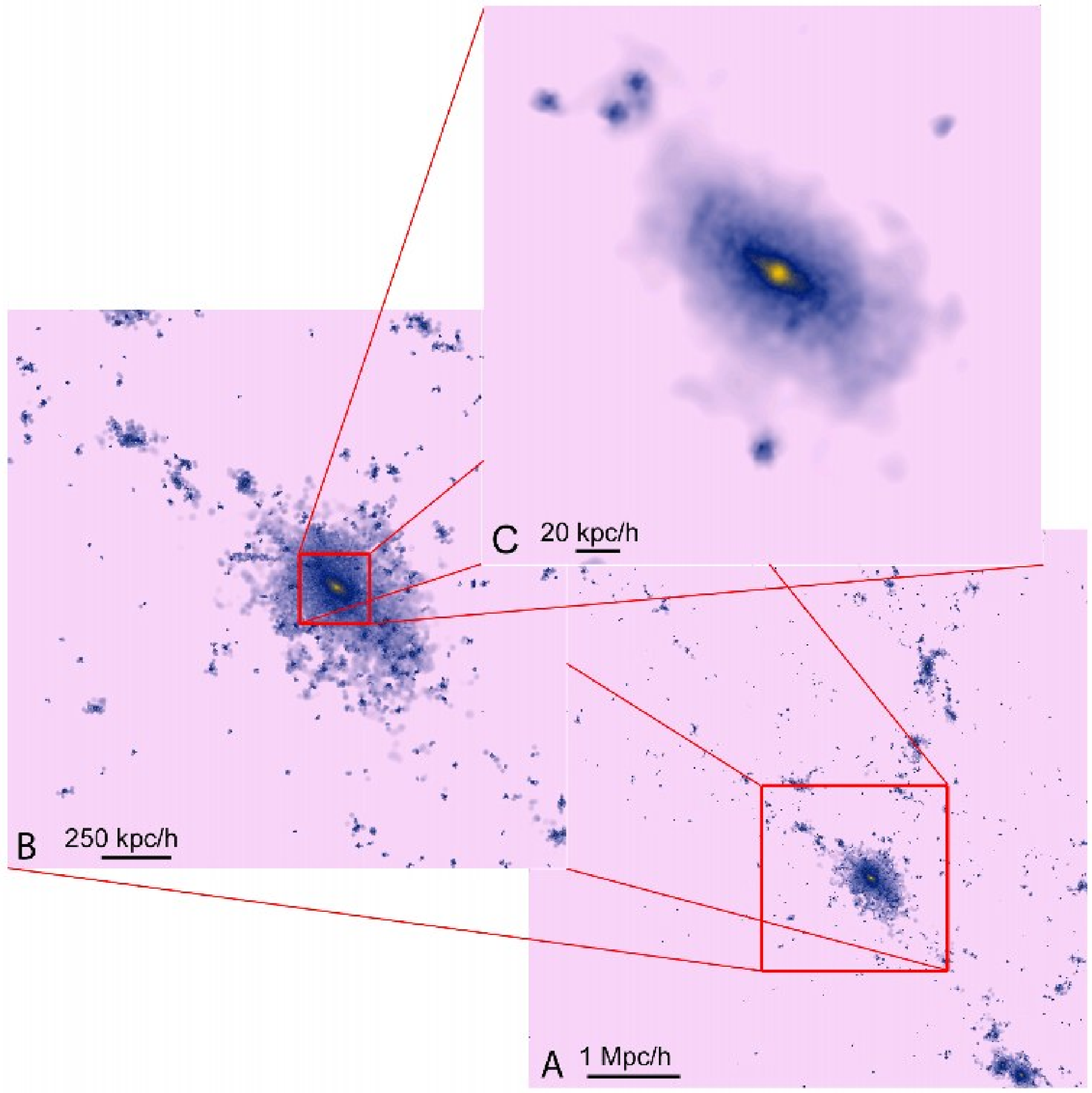}}
\end{center}
\caption{Stellar density in a three level zoom into the same regions
shown in Figure~\ref{fig:D6zoomsgas}, and which is marked by a white
rectangle in the $z=3$ panel of Fig~\ref{fig:D6slices}. In the most
zoomed-in panel C, a stellar disk and a small bulge component can be
seen for the central object.}
\label{fig:D6zoomsstars}
\end{figure*}

\subsection{Simulation runs}
Cosmological simulations which include quasar formation must model
sufficiently large volumes to sample a representative part of the universe,
but also have high enough resolution to model the full hydrodynamics. This is
a substantial challenge, given that the brightest quasars at $z\sim 6$ have a
low space density and are believed to reside in fairly massive dark matter
halos of mass $M\sim 10^{11}-10^{12} \Msun$, or even larger.  
At redshifts $z=2-3$, quasars
have a much larger space density, comparable to $L^{*}$ galaxies at $z=0$.

The strategy we choose here is model the universe with a periodic box of
moderate size $33.75\,h^{-1}{\rm Mpc}$ that is homogeneously sampled with
particles, and which we simulate with $2\times 486^3$ particles, one of the
{\it highest resolutions} so far achieved in a full cosmological
hydrodynamical calculation of galaxy formation. In this paper, we refer to
this largest simulation among our simulation set as the {\it BHCosmo} run.  We
will also compare it with two additional simulations which differ in mass and
spatial resolution, and/or box size, to test for resolution effects.

The fundamental numerical parameters of our simulation runs are listed in
Table~\ref{tab:simul}, where $N_p$ is the number of dark matter and gas
particles, $m_{\rm DM}$ and $m_{\rm gas}$ are their masses (initial mass in
the case of the gas). Finally, $\epsilon $ gives the comoving gravitational
softening length, and $z_{\rm end}$ the final redshift of the simulation.

For the physical problem at hand we prefer relatively high resolution in order
to capture the physics in high density regions appropriately, but we also need
a large volume to study the growth of deep gravitational potentials. Our
choice of $33.75\,h^{-1}{\rm Mpc}$ represents a compromise in this respect.
While this box-size is too small to be evolved to redshifts lower than
$z\simeq 1$ (otherwise the fundamental mode would become non-linear), it is
sufficiently large to provide a representative model for $L_\star$-objects at
higher redshift, even though very rare systems in the exponential tail of the
mass function will not be sampled well.  We also note that the choice of
box-size and particle number in the {\it BHCosmo} run is such that the
physical resolution at $z\sim 6$ is comparable to that in some of our previous
works on galaxy mergers, namely runs which used only $N\simeq 10000$ particles
for each galaxy. In this prior work, we have shown that despite the low
resolution the results for the black hole mass growth agreed well with those
obtained in runs with 128 times higher resolution, and can therefore be
considered converged with respect to this quantity. This overlap in
resolution between our cosmological runs and our previous work on isolated
galaxy mergers gives us confidence that the results of our cosmological runs
are not dominated by resolution effects, although it is clear that this needs
to be tested separately.  We remark that as part of our previous work on
mergers we have run a suite of several hundred galaxy merger simulations
\citep{Robertson2006a}, varying all the parameters describing star formation
and feedback from supernovae and black hole growth and accretion, besides
carrying out numerical resolution studies. The galaxy merger simulations are
clearly much better suited for investigating the full parameter space of our
model, while for the cosmological runs we have to restrict ourselves to our
default model owing to their much larger computational cost.

In addition to the above considerations, the choice of box-size in our new
simulations is also motivated by the set of simulations presented in
\citet{Springel2003b}. In fact, the {\it BHCosmo} run would be called `D6' in
their naming scheme. Being able to directly refer to their runs simplifies the
comparison of the physical properties of simulations with and without black
holes, e.g.~with respect to the star formation history.



\section{Results}

\subsection{Visualization of the structure and black hole growth}
In Figure~\ref{fig:D6slices}, we show slices through the {\it BHCosmo}
simulation at a range of redshifts in order to visualize the evolution of the
baryonic density field and the growth of black holes.  The slice has a
thickness of $5\,h^{-1} {\rm Mpc}$ and shows the full box of size $33.75\,
h^{-1}{\rm Mpc}$ on a side. In each panel, the projected gas density field is
color-coded according to the gas temperature, with the brightness of each
pixel being proportional to the logarithm of the gas surface density. Circles
of different size are drawn to mark the locations of BHs of different mass, as
labelled.

The images show that black holes emerge in halos starting at high redshift (as
early as $z\sim 12$) and subsequently grow by gas accretion, driven by gas
inflows that accompany the hierarchical build-up of ever larger halos through
merging. As the simulation evolves, the number of black holes rapidly
increases and larger halos host increasingly more massive black holes.  We
note that the particular slice of the box shown in Figure~\ref{fig:D6slices}
does not contain the largest black hole in the simulation's volume, which turns
out to be located in the highest density region in the simulation.  We will
discuss this region separately in later sections.

By plotting the density field color-coded by temperature we can also see
traces of heating effects from strong gas outflows, which are caused by black
hole feedback and galactic winds from star formation. In our numerical mode,
quasar-driven outflows occur once a central BH gas grown so much that its
energy feedback in accretion phases is able to transfer sufficient energy to
the remaining gas to unbind and expel part of it from the galaxy potential as
a wind. This then terminates further strong growth of the BH, which is key
factor in establishing a self-regulated nature of the growth of black holes in
our model, as we discussed in detail in our previous work on galaxy merger
simulations.

To illustrate how well the mass resolution of the simulation captures details
of galaxy formation sites, we zoom in onto a region of $6\,h^{-1}{\rm Mpc}$ on
a side at $z=3$ (as indicated by the box drawn in Fig~\ref{fig:D6slices}),
showing the gas density in Figure~\ref{fig:D6zoomsgas}, and the stellar
density of the same region in Figure~\ref{fig:D6zoomsstars}. The middle
panels (labeled B) in both figures show a second zoom-level into a region of
$2\,h^{-1}{\rm Mpc}$ on a side (indicated by the square in the A panels).
Finally, the top panels show a further enlargement by a factor of 8.

In accordance with our seeding procedure, black holes are located in the
highest density regions. The more massive ones are found within the largest
halos, which have also undergone more prominent star formation. Such a
correspondence is expected in our model since large-scale gas inflows into the
centers of halos lead both to star formation (and starbursts) as well as to
nuclear black hole growth. In the highest level zoom of
Figures~\ref{fig:D6zoomsgas} and \ref{fig:D6zoomsstars}, the central galaxy,
with an extent of $\sim50h^{-1} {\rm kpc}$ (comoving), has a very rough 
disk-like morphology with a central stellar bulge.
Nevertheless, it is clear that our cosmological simulations in general still
have too low resolution for properly resolving galaxy morphologies. We also
note that producing disk galaxies with the right size and abundance in
cosmological hydrodynamical simulations is an essentially unsolved problem,
and the outlook for obtaining a solution to this long-standing challenge has
only slightly improved by recent works on disk galaxy formation
\citep{Robertson2004, Governato2007, Okamoto2007}.

\subsection{The evolution of the global black hole mass density}
The black hole mass density at the present epoch is estimated from direct
measurements of black hole masses in local galaxies (to establish, e.g., the
$M_{\rm BH}-M_\star$ relationship), combined with a suitable integration over
the galaxy luminosity function ~\citep{Fabian1999, YuTremaine2002,
Marconi2004}. The local value thus obtained matches that of the total relic BH
mass density estimated from a time integration of the luminosity output of
active galactic nuclei and quasars at X-ray and optical wavelengths
\citep{Soltan1982, Marconi2004, Shankar2004}, or in a less model-dependent
manner from an empirical determination of the bolometric quasar
luminosity function \citep{Hopkins2007b}.

\begin{figure}
\hspace*{-0.5cm}\vspace*{-1.0cm}\\
 \resizebox{9.0cm}{!}{\includegraphics{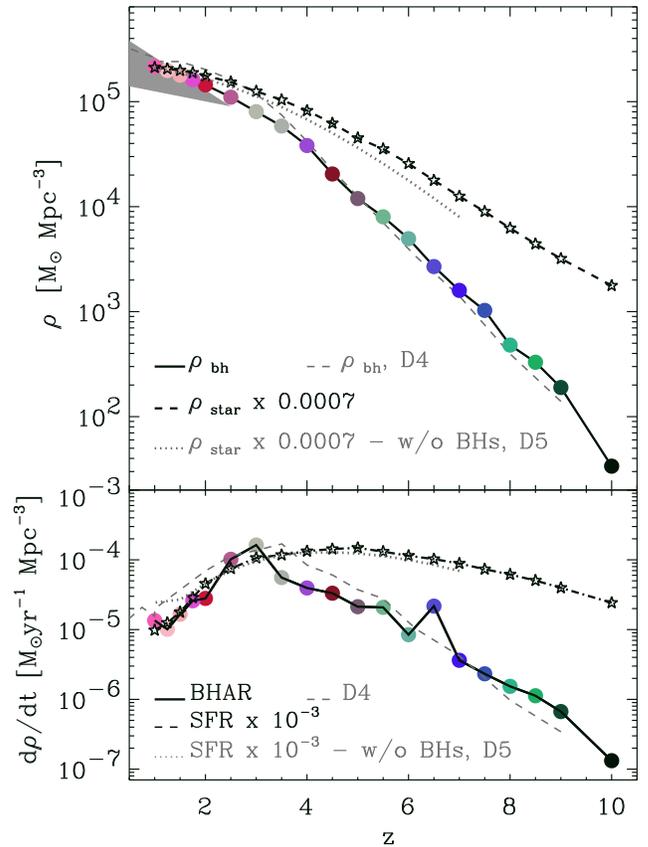}}
\caption{{\it Top panel}: The global black hole mass density evolution in the
{\it BHCosmo}/D5 run, shown by a solid line. The the star symbols and the
dashed line give the corresponding stellar mass density evolution, multiplied
by 0.0007 for easier comparison.  The grey dotted line shows the stellar mass
density in the D5 simulation which did not include black hole accretion and
associated quasar feedback ~\citep{Springel2003a}. The thin dashed line shows
the results from the lower resolution box, D4 described in
Table~\ref{tab:simul}. Different colors simply indicate the different
redshifts consistent with the scheme used in other figures.  The shaded grey
triangle indicates observational constraints taken from the literature
\citep{Marconi2004, Shankar2004}.  {\it Bottom Panel}: The global history of
the black hole accretion rate (solid line) and star formation rate (dot-dashed
line with stellar symbols) densities. The SFR is rescaled by $10^{-3}$ for
graphical clarity. In addition, we show the SFR history in the D5 simulations
without black holes (grey dotted line). Most of the black hole and stellar
mass is assembled by $z\sim 2-3$, but the peak in the BHAR density function is
far more pronounced than that of the SFR density. }
    \label{fig:D6acc_den}
\end{figure}

Figure~\ref{fig:D6acc_den} shows our simulation prediction for the global
density $\rho_{\rm BH}$ and its evolution with redshift ({\it thick black
  line}). We find that the normalization of the black hole mass density is in
agreement with the observational estimate of $\rho_{\rm BH} (0) =
4.6^{+1.9}_{-1.4} \times 10^{5} h_{0.7}^2 \Msun \Mpc^{-3}$ of
\citet{Marconi2004} and its extrapolation to $z\sim 3$, derived by exploiting
hard X-ray and optically selected AGNs and quasars.  The grey area in
Figure~\ref{fig:D6acc_den} shows the region delimited by observational
constraints examined in detail in the literature \citep[e.g][]{Salucci1999,
  Marconi2004, Shankar2004}. The agreement we see here is very reassuring.  It
shows that our black hole accretion and feedback model in the {\it BHCosmo}
run is adequate and provides a realistic account for the dominant mode of
global black hole growth in our universe.

The black hole mass density evolves rapidly at high redshift, increasing by
four orders of magnitude between $z \sim 10$ to $z \sim 3$. While below this
redshift, there is some further growth, it only accounts for roughly a
doubling of the BH density down to $z= 1$. This is corroborated by the bottom
panel of Figure~\ref{fig:D6acc_den}, where we plot the history of the global
black hole accretion rate (BHAR) density in our {\it BHCosmo} run. It is
evident that the black hole mass growth occurs mainly by accretion above $z
\sim 3$.  Indeed, the BHAR density rises steadily at early times to a peak at
$z \sim 3$, where it reaches a value $ \sim 10^{-4} \Msun \yr^{-1} \Mpc^{-3}$.
Below this redshift, it drops rather rapidly, becoming more than an order of
magnitude lower by $z= 1$.  We note that the 'spike' seen in the BHAR density
at $z \simeq 6.5$ is caused by a single object in the box, namely the rapid
formation of the most massive black hole in the simulation at this epoch. We
will discuss the history of the corresponding halo and its embedded black hole
in more detail in \S 5.

\begin{figure}
\hspace{-1cm}
      \resizebox{10.cm}{!}{\includegraphics[width=4.in]{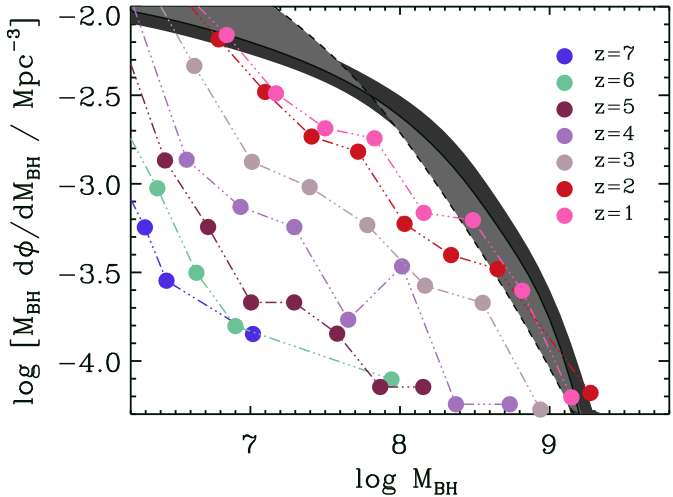}}
    \caption{The evolution of the black hole mass function in the {\it
BHCosmo} run. The different colors indicate different redshifts as
labeled in the figure. For comparison, the dark grey shaded region
shows the black hole mass function derived from local galaxies
\citep{Marconi2004}. The light gray area adds the constraint from the
integration of the hard X-ray luminosity function \citep{Shankar2004,
Merloni2004, Marconi2004}.  The largest uncertainties are at the high
and low mass end. The simulation results are in good agreement with
the observed mass function at the high mass end, and in reasonable
agreement at intermediate masses.}
    \label{fig:bhmf}
\end{figure}

\begin{figure}
\hspace{-1cm}
      \resizebox{10cm}{!}
{\includegraphics[width=4.in]{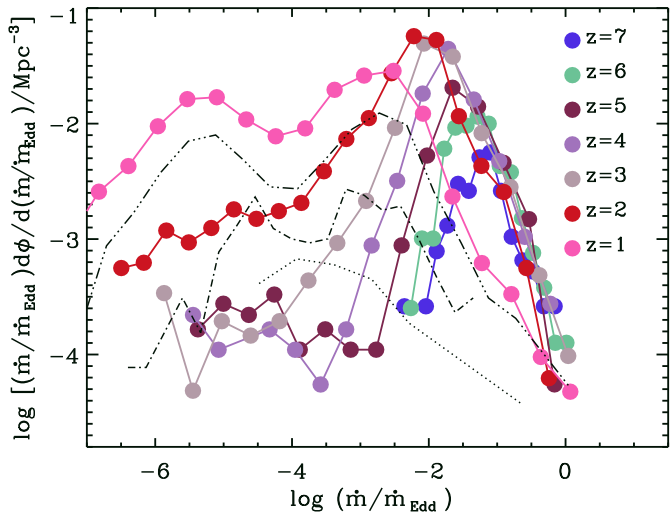}}
\caption{The time evolution of the accretion rate distribution as a
  function of the Eddington ratio, for the {\it BHCosmo} run. The
  different colors denote our measurements at different redshifts, as
  indicated in the legend. For the $z=1$ distribution function, we
  separately show three components corresponding to different regions
  of the black hole mass function.  In particular, the dotted,
  dot-dashed, and dot-dot-dashed lines give the separate contributions
  from the three different mass bins $M_{\rm BH} > 10^{8} \Msun$,
  $10^7 < M_{\rm BH}/\Msun \le 10^{8}$ and $10^6 \leq M_{\rm BH}/\Msun
  <10^{7}$, respectively. 
}
    \label{fig:bhaf}
\end{figure}

\subsection{Black hole mass function and accretion rate function}
\label{sec:massfn}
In Figure~\ref{fig:bhmf}, we plot the black hole mass function at a
number of different redshifts. We find that the final black hole mass function
in our simulation (for $z=1$) is is quite good agreement with the one measured
locally, especially on the high-mass side. The $z=0$ constraint is indicated
by the {\em dark grey area} taken from the compilation of \citet{Marconi2004},
which is based on a combination of different observational data
\citep{kochanek2001, Nakamura2003, Bernardi2003, Sheth2003}. There is a small
deficit in our model at intermediate BH masses, but note that this will be
filled in at least partly by the expected residual growth from $z=1$ to $z=0$.
The {\em light grey area} adds an additional constraint for the contribution
of relic AGN, derived from an integration of the the hard X-ray luminosity
function \citep{Shankar2004}.  (Note that in this latter case the
normalization of the mass function depends on the value assumed for the
radiative efficiency when converting from the luminosity function to the mass
function.)
 
\begin{figure*}
\begin{center}
\resizebox{8.5cm}{!}{\includegraphics{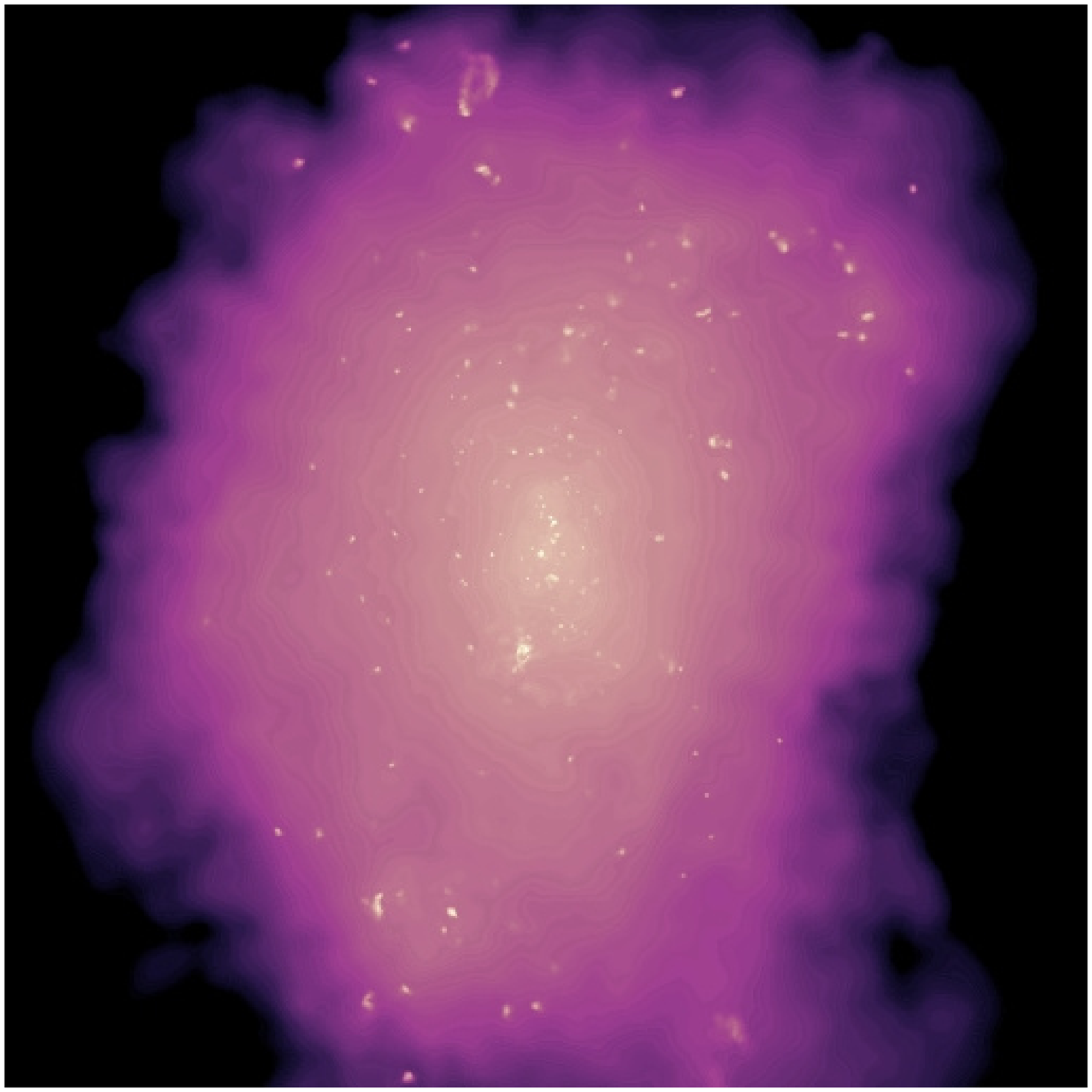}} %
\resizebox{8.5cm}{!}{\includegraphics{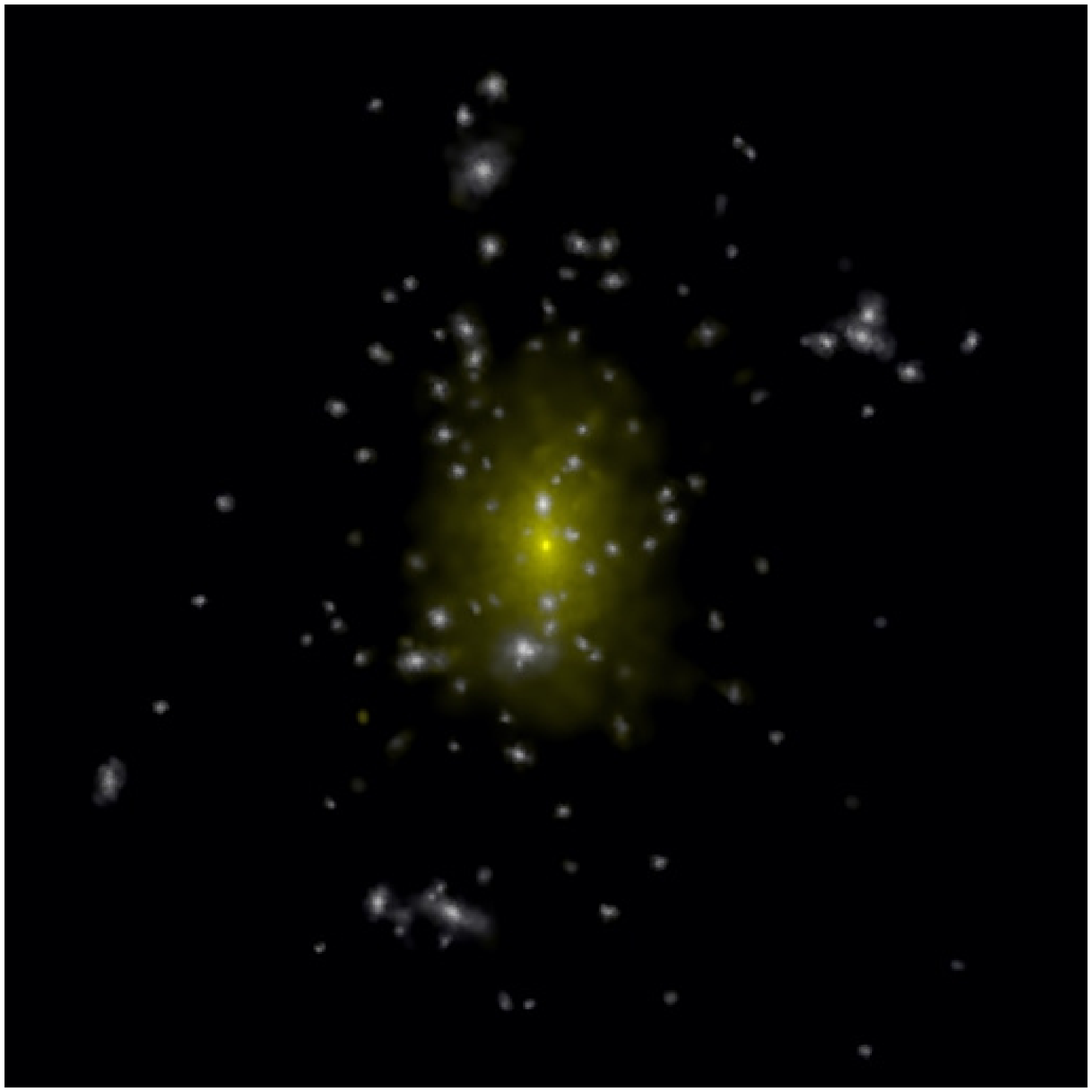}}
\end{center}
    \caption{An illustration of a large group in the {\it BHCosmo} run at
$z=1$.  {\it Left panel:} the gas density distribution. {\it Right
panel:} the stellar distribution. We have run a sub-group finder to
identify all the systems (galaxies) within each halo and analyze their
properties accordingly. The stellar system shown in yellow on the
right hand side is the main galaxy within this large group, while the
satellite galaxies are shown in grey. The images are 400 kpc on a side.
    \label{fig:subgroups}}
\end{figure*}

\begin{figure*}
\begin{center}
\resizebox{18.0cm}{!}{\includegraphics{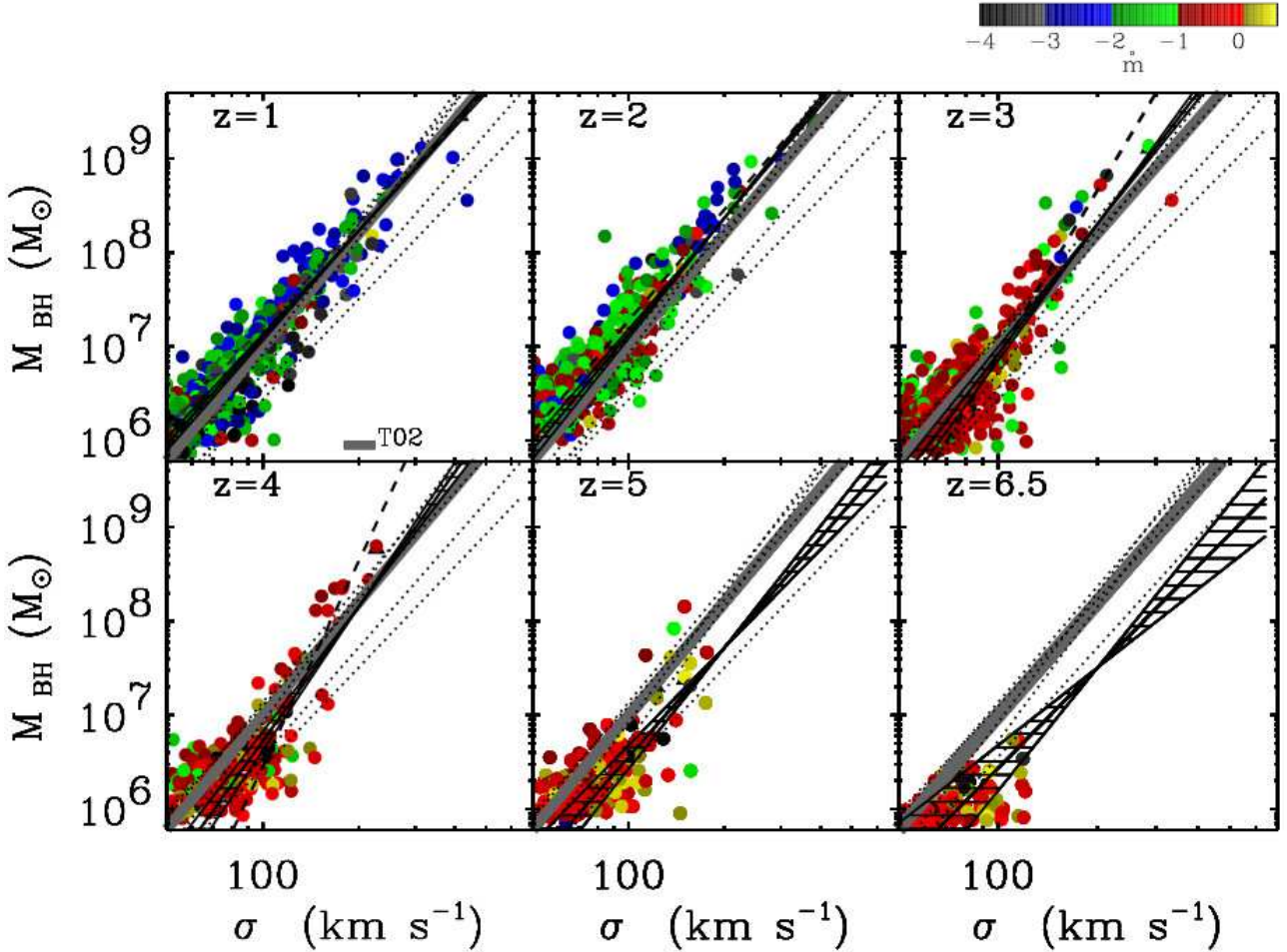}}
\end{center}
    \caption{The evolution of the $M_{\rm BH} - \sigma$ relation in the {\it
BHCosmo} simulation. The masses of BHs and the projected stellar velocity
dispersions within the half mass radius ($R_e$) have been measured in our
simulated galaxies and are plotted at $z=1$, $2$, $3$, $4$, $5$, and $6.5$. We
compared our measurements at all redshifts with the best-fit to the local
$M_{\rm BH}- \sigma$ relation of \citet[][hereafter T02]{Tremaine2002}, which
is shown as a thick gray line. Linear regression fits to our simulated BHs are
shown by solid lines at each redshift, with $1-\sigma$ errors indicated by the
hatched regions. For ease of comparison, the dotted lines in each panel show
the best fit relations at all redshifts. The points are color-coded
according to their accretion rates in units of Eddington, as indicated in the
color bar at the top right hand corner of the figure. 
}
\label{fig:mbhsigma}
\end{figure*}

\begin{figure*}
\begin{center}
  \resizebox{18.0cm}{!}{\includegraphics{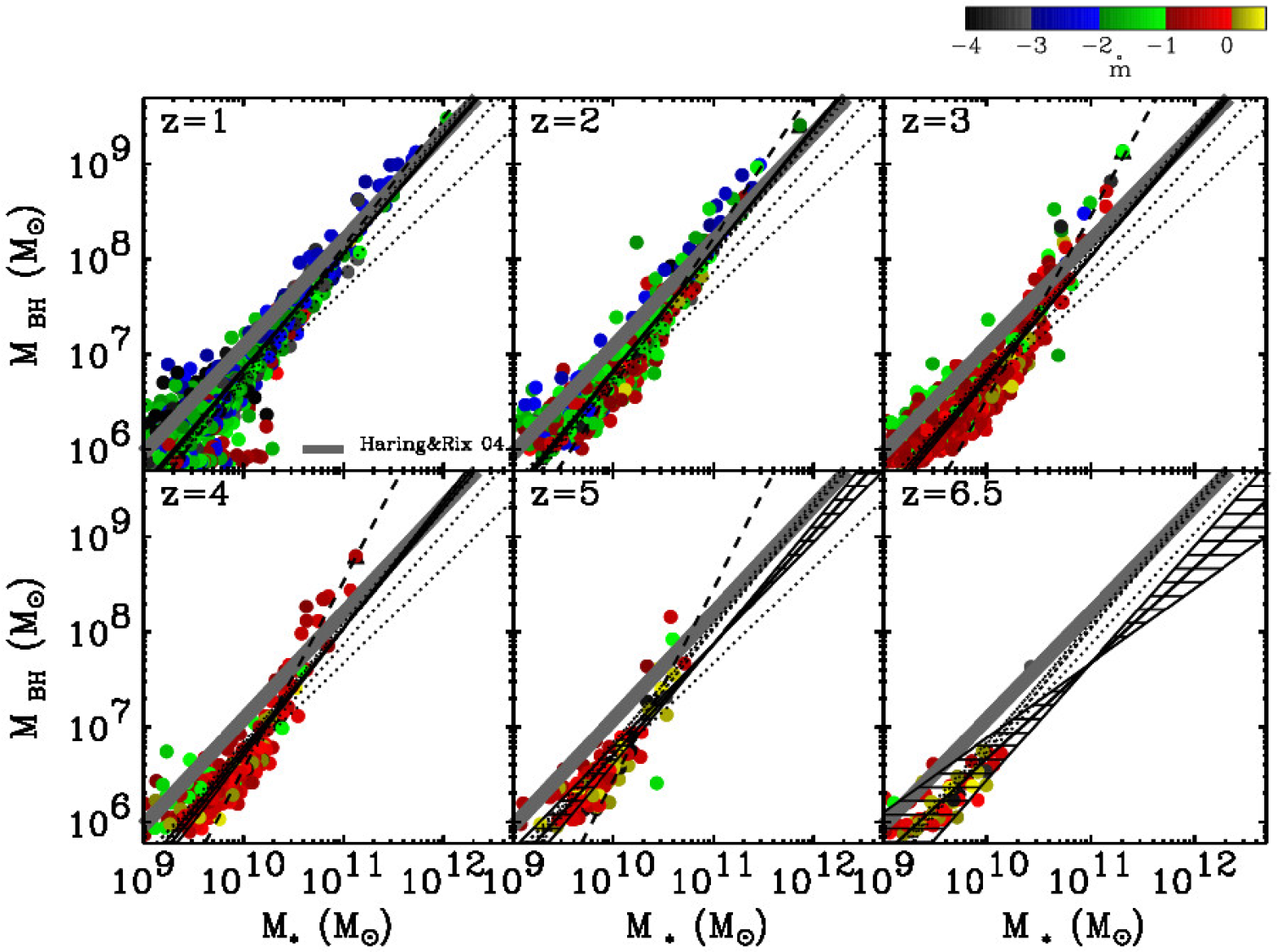}}
\end{center}
    \caption{The evolution of the $M_{\rm BH} - M_{*}$ relation in the
{\it BHCosmo} simulation. The masses of BHs and the corresponding
stellar mass have been measured in our simulated galaxies and are
plotted at $z=1$, $2$, $3$, $4$, $5$, and $6.5$. They are compared
with the best-fit power law to the local $M_{\rm BH}- M_{*}$ relation
by \citet{Haring2004}, shown by the thick gray line. Our fits to
the simulation results at each redshift are shown as solid black
lines. The relation is tight with small scatter ($1-\sigma$
uncertainty ranges are plotted as hatched regions but are hardly
visible at low redshift). As in Figure~\ref{fig:mbhsigma}, the points
are color coded by accretion rate. The dotted lines show in each panel
the best fit relation at the other redshifts. 
    \label{fig:mbhmstar}}
\end{figure*}

As expected in a hierarchical formation scenario, the high mass end of the
mass function shifts to larger masses with redshift. However, it is
interesting that the mass function for high masses grows rather rapidly at
early times relative to the low mass end.  On the other hand, for redshifts $z
\lesssim 2$, the high mass end is virtually fully assembled while the BH mass
function continues to grow for low to intermediate masses, leading effectively
to a steepening of the mass function for $z \simlt 3$.  This appears
consistent with the emerging observational picture of the evolution of the
supermassive black hole population according to which massive BHs ($M_{\rm BH}
\simgt 10^9 \Msun$) are assembled early and are then likely undergoing
comparatively passive evolution in the centers of large spheroids
\citep{Shields2003, Vestergaard2004, Adelberger2005}. More generally, this
phenomenon is described by the idea of an `anti-hierarchical' black hole
growth, or equivalently a `downsizing' of black hole activity
\citep{Merloni2004, Marconi2004}, which is derived from constraints on the
accretion history from X-ray luminosity functions.


Finally, in Figure~\ref{fig:bhaf}, we show the evolution of the corresponding
accretion rate distribution function for the black holes in our simulation,
expressed in units of the Eddington rate. This function is strongly peaked at
a few percent below the critical Eddington value, with most black holes
accreting at $10^{-2} \lesssim \mdot/\mdot_{\rm Edd} \lesssim 1 $ at redshifts
$z\simgt 6$. The distribution becomes wider and develops a small amplitude
tail at low Eddington rates for $ 6 < z < 3$.  Below $z \sim 3$ the peak of
the accretion rate function shifts to $\mdot/\mdot_{\rm Edd} \sim 10^{-2} -
10^{-3}$ and an increasingly large population of sources is present accreting
down to $10^{-6}$ Eddington. At $z\sim 1$ the distribution function is
extremely wide and dominated by sub-Eddington (to severely sub-Eddington)
sources, in terms of number. While hence the number of sources accreting at
sub-Eddington rates increases sharply with decreasing redshift, the
quasar-like population, i.e.~the sources accreting close to Eddington,
decreases with decreasing redshift, with a maximum at $z\sim 3$.  For $z\sim
1$, we also include in Figure~\ref{fig:bhaf} measurements of the separate
contributions to the accretion rate function from three different black hole
mass ranges.  The dotted line gives the Eddington ratio distribution for
$M_{\rm BH} > 10^{8} \Msun$, and the dash-dotted and dot-dot-dashed lines
are for masses $10^7 < M_{\rm BH}/\Msun \le 10^{8}$ and $10^6 \leq M_{\rm
BH}/\Msun <10^{7}$, respectively.

Taken together, the evolution of the black hole mass and accretion rate
functions implies that most massive black holes assemble early and do so at
close to their critical rate. At low redshift, progressively lower
luminosities and lower mass systems start dominating the black hole activity.
This is in accord with recent results from studies of the hard X-ray
luminosity function of quasars and AGN, e.g.~typical Seyfert-like objects with
$M_{\rm BH} \sim 10^7 - 10^8 \Msun$ accreting at a few percent of the
Eddington rate~\citep[e.g.][]{Ueda2003, Hasinger2005}.

\subsection{Comparison of the histories of black hole growth and star formation}In Figure~\ref{fig:D6acc_den}, we have already shown and discussed the
evolution of the black hole mass and accretion rate densities. In the
same Figure, we show corresponding results for the evolution of the
stellar mass density, $\rho_{*}$, and the star formation rate (SFR)
density, with their normalizations rescaled by factors of $7\times
10^{-3}$ and $10^{-3}$, respectively, for plotting purposes. For
comparison, we also include results for the D5 simulation of
\citet{Springel2003b} which did not include black holes and any
associated accretion or feedback processes (dotted grey lines).

We can see that $\rho_{*}$ far exceeds $\rho_{\rm BH}$ at all redshifts, with
$\rho_{*}$ evolving less strongly with redshift than $\rho_{\rm BH}$ for $z
\simgt 3$. At early times, $\rho_{\rm BH}$ rises more rapidly than the star
formation density, while it tracks its evolution below this redshift. If we
parameterize the ratio of $\rho_{\rm BH}/ \rho_{*}$ by an evolutionary factor
$(1+z)^{\alpha}$ we find that its evolution is approximately given by
\begin{equation}
\rho_{\rm BH}/ \rho_{*} \sim \left\{ \begin{array}{ll} \phi_{3} (1+z)^{-3} & \mbox{if $z \ge
3.5$} \nonumber \\ & \\ \phi_1 [(1+z)]^{-0.6} & \mbox{if $z <
3.5$}
\end{array} \right. , \label{eq:rho}
\end{equation}
with $\phi_{3} = 5 \times 10^{-2}$ and
$\phi_{1}=2.2\times10^{-3}$. Accordingly, up to $z\sim 3$, the evolution of
the star formation rate density is considerably shallower than that of the
black hole accretion rate density. Below this redshift, the BHAR and SFR
densities closely track each other.  As a result, the BHAR density has much
more pronounced peak, which we find to lie at slightly lower redshift than the
peak of the SFR. Parameterizing the evolution of the BHAR and SFR density
ratio with a power law in $(1+z)$ leads to
\begin{equation}
\dot{\rho_{BH}} / \dot{\rho}_{*} \sim\left\{ \begin{array}{ll} \dot{\rho}_{3} (1+z)^{-4} & \mbox{if $z \ge
3$} \nonumber \\ & \\ \dot{\rho}_1 [(1+z)]^{-0.1} & \mbox{if $z <
3$}
\end{array} \right. , \label{eq:drho}
\end{equation}
with $\dot{\phi}_{3} = 0.2$ and $\dot{\phi}_{1} = 10^{-3}$. 

Note that Equations~(\ref{eq:rho}) and (\ref{eq:drho}) are only meant to
provide approximate scalings for our results from the simulations. The
important point we want to emphasize is that our results imply a different and
much stronger evolution of the black hole mass and accretion rate densities at
high redshift relative to the stellar density and star formation rate density.
The black hole mass density tends to be assembled later than the stellar mass,
despite the growth of (a small number of) very massive BHs already at high
redshift. However, for $z \sim 3$ and below, our models predict that the black
hole mass and stellar mass densities, as well as the BHAR and SFR densities,
closely track each other allowing only for small amounts of relative evolution
between the two. 

Another important point from Figure~\ref{fig:D6acc_den} is that the feedback
we associate with black hole accretion does not significantly affect the
global assembly of stellar mass. The peak of the SFR density is unaffected by
the inclusion of BH feedback, but the drop in SFR density ($z < 3$) is
slightly more abrupt in the simulations with black holes. This effect becomes
more pronounced at $z\sim 1$, the final redshift for our simulation.  At still
lower redshift, we expect that BH feedback will become important in regulating
the cooling and star formation in very massive halos. This is then ascribed
not to quasar growth but rather to a `radio mode'. We explore this different
feedback channel in a companion paper by \citet{Sijacki2007}.

\section{The $M_{\rm BH} - \sigma$ and $M_{\rm BH} - M_{*}$ relations}
\label{msigmarel}

\subsection{Identification of groups and subgroups}
As a prerequisite for being able to study correlations between black
hole and host galaxy properties in our simulation we first need to
apply a suitable group finding algorithm that reliably identifies the
stellar mass associated with the different galaxies. Note that
especially the more massive halos identified by our basic
friends-of-friends grouping algorithm used for finding virialized
objects often contain a number of galaxies. This is illustrated in
Figure~\ref{fig:subgroups}, where we show a large cluster-sized group
selected from the $z=1$ output of the {\it BHCosmo} simulation. The
panel on the left shows the gas density distribution in this large
group, while the panel on the right hand side displays the stellar
distribution. It is evident that the group contains several,
gravitationally bound galaxies.

Our sub-group finder identifies all the galaxies within each group.  Our
method to identify galaxies within a given group is based on a variant of the
{\small SUBFIND} algorithm \citep{Springel2001}.  We first compute an
adaptively smoothed baryonic density field for all stars and gas particles,
allowing us to robustly identify centers of individual galaxies.  We find the
extent of these galaxies by processing the gas and star particles in the order
of declining density, adding particles one by one to the galaxies attached to
the galaxy to which its nearest denser neighbor already belongs~\citep[see
also][]{Nagamine2004}. Note we are interested in the stellar and gas content
of galaxies and associating the gaseous component to galaxies, as they
typically they contain very dense star-forming gas, makes the method very
robust in finding galaxies. This allows us to compute physical properties
such as stellar mass, star formation rate, stellar velocity dispersion,
metallicity, black hole mass and BH accretion rate separately for each
galaxy. As an illustration, Figure~\ref{fig:subgroups} shows the stellar
distribution associated with the largest galaxy in the group (which also hosts
the most massive BH in the group) in yellow.

For each galaxy/subgroup that contains stars and a black hole, we calculate the
projected (spherically averaged) half-mass effective radius, $R_e$, and the
mass-weighted stellar velocity dispersion $\sigma$ within $R_e$.  Note however
that in order to make the measurements of $\sigma$ and $R_e$ somewhat accurate
we only consider those objects that contain more than 100 stars particles
within $R_e$.  This determination of $\sigma$ closely resembles the procedure
for measuring the velocity dispersion from observational data
\citep{Gebhardt2000}, allowing for a direct comparison.

\subsection{The predicted $M_{\rm BH} - \sigma$ and $M_{\rm BH} -M_{*}$
  relationships and their evolution} Figures~\ref{fig:mbhsigma} and
\ref{fig:mbhmstar} plot the $M_{\rm BH} - \sigma$ and $M_{\rm BH} -M_{*}$
relations, respectively, for our simulated galaxies at redshifts $z=1$, $2$,
$3$, $4$, $5$ and $6.5$ (from top left to bottom right). Each measurement is
color-coded according to the accretion rate of the corresponding black
hole. We find a strong power-law correlation between both the velocity
dispersion and the stellar mass with the black hole mass. Furthermore, at
$z=1$, these correlations agree very well with the ones observed at the
present epoch~\citep{Ferrarese2000, Gebhardt2000, Tremaine2002, Haring2004} 
over a large dynamic range. This is a remarkable confirmation of the basic
merger-driven scenario for self-regulated BH growth that we previously
explored in isolated high-resolution galaxy merger simulations. Based on the
same model, and without any fine-tuning or change of the model parameters, we
here obtain a good match to the observed $M_{\rm BH} - \sigma$ relation
directly from simulations that start from cosmological initial conditions and
self-consistently account for the hierarchical build up galaxies in a
$\Lambda$CDM universe.

\subsubsection{The evolution of $M_{\rm BH} -\sigma$}
At higher redshifts, and in particular for $z > 3$, our results for the
$M_{\rm BH} - \sigma$ relation shows some degree of evolution relative to the
local relations. To compare with the local  data as a function of time,
 we
fit the $M_{\rm BH}-\sigma$ relation at each redshift 
with a simple power-law of the form \beq \log \left(\frac{M_{\rm
      BH}}{\Msun}\right) = a\, \log \left(\frac{\sigma}{200 \kmps}\right) + b.
\label{eqn:msigmafit}
\eeq The best-fit relations thus obtained are shown in
Figure~\ref{fig:mbhsigma} as solid black lines, with the dispersion indicated
by hatched regions. The dotted lines are the best-fit relations for all
redshifts combined. We compare with the observed relation as determined by
T02, which is described by a slope $a=4.02$, a normalization $b=8.2$ and a
dispersion $\Delta \sim 0.25-0.30$ (grey thick line in
Fig.~\ref{fig:mbhsigma}).

\begin{table}
\begin{center}
\caption{Parameters of best-fit $M_{\rm BH} -\sigma$ relations}
\label{tab:msigma}
\begin{tabular}{ccccc}
\hline\hline\\
z  &  slope (a) & normalization, (b) & scatter $\Delta$ & slope (a$_{s}$)\\
 \hline\\
1 ....... &  $3.95\pm 0.10$ & $8.29\pm0.03$ & 0.10 & 3.9  \\
2 ....... &  $4.01\pm 0.15$ & $8.42\pm0.04$ & 0.16 & 4.1  \\
3 ....... &  $4.21\pm 0.22$ & $8.32\pm0.07$ & 0.23 & 5.2 \\
4 ....... &  $4.54\pm 0.35$ & $8.17\pm0.10$ & 0.36 & 6.7 \\
5 ....... &  $3.37\pm 0.45$ & $7.70\pm0.13$ & 0.47 &  --   \\
6.5 ..... &  $3.26\pm 0.85$ & $7.50\pm0.25$ & 0.89 &  --  \\
\hline\\
\end{tabular}
\end{center}
\vspace{-1cm}
\end{table}

The constants $a$ and $b$ for our best-fit relations and their dispersions are
tabulated in Table~\ref{tab:msigma} for different redshifts.  Note that we
here do not attempt to assess the statistical significance of the correlations
in detail as the sources of systematic errors in the numerical measurements of
$\sigma$ and $M_{\rm BH}$ cannot be easily quantified in the simulations.  In
particular, the cosmological simulations cannot determine the morphological
properties of galaxies and therefore do not provide a direct measure
of spheroid masses or their velocity dispersions.  Our
fitting procedure is merely intended to provide a first characterization of
the overall evolution of the slope and normalization of the relations in the
simulation model.  As Figure~\ref{fig:mbhsigma} and Table~\ref{tab:msigma}
indicate, the $M_{\rm BH}-\sigma$ relation predicted from our simulation is
consistent with a slope $\sim 4$ at low redshift, as observed. At $z=3-4$, the
slope appears to be slightly steeper and at $z=5-6$ slightly shallower, but the
small number of systems at $z \sim 6$ makes the latter trend uncertain.

Inspection of Figure~\ref{fig:mbhsigma} shows a qualitative trend whereby the
larger systems with high $\sigma \simgt 150 \kmps$ (which are also those that
are best resolved in our simulation) appear to populate the high mass end of
the $M_{\rm BH}-\sigma$ relation already at $z > 2-3$.  The lower mass end of
the $M_{\rm BH}-\sigma$ relation is then increasingly filled in towards $z\sim
1$. To illustrate this trend more explicitly, we perform a fit of the relation
using only those systems with $\sigma \ge 150\kmps$; this is shown by the
dashed line in Figure~\ref{fig:mbhsigma}. The slope $a_{s}$ obtained for these
`high' $\sigma$ black holes is typically steeper than for the full relation
(our measured values for $a_{s}$ are listed in Table~\ref{tab:msigma}). At a
fixed and relatively high $\sigma$, the mean $M_{\rm BH}$ is larger at $z\sim
3-4$ than at $z\sim 1$.  This trend suggests that black hole growth predates
the final growth of the spheroid potential at scales $\sigma \sim 170\,{\rm
km\,s^{-1}}$, which is consistent with the recent measurements of the $M_{\rm
BH}-\sigma$ relation in a sample of Seyferts at $z\sim 0.35$ by
\citet{Woo2006}. This relatively more efficient black hole growth at a fixed
$\sigma$ at high redshift may thus be responsible for a mild change in the
high mass end of the $M_{\rm BH}-\sigma$ relation. In the remainder of this
section we will discuss possible additional dependences in the $M_{\rm
BH}-\sigma$ \citep{Hopkins2007a} relation which may produces a small change in
the slope of the relation with time.

Note also that at $z \ge 4$ black holes are more likely to accrete close to
their critical Eddington rates, as we showed earlier. In our models, the
$M_{\rm BH}-\sigma$ relation is a natural consequence of the self-regulated
growth of black holes \citep{DiMatteo2005}, where a black hole grows until its
released energy is sufficient to expel the gas in its surroundings in a quasar
driven wind, which then terminates nuclear accretion. For this reason, we
expect the relation to show more scatter at times when most systems are still
actively growing and accreting close to their Eddington values (see
Table~\ref{tab:msigma}). This is expected as the primary path for assembly BH
mass is via accretion during major mergers so that the relations converge and
get increasingly tighter as galaxies undergo major mergers and continue to
merge.

\begin{figure}
\hspace*{-0.4cm}
\resizebox{8.5cm}{!}{\includegraphics{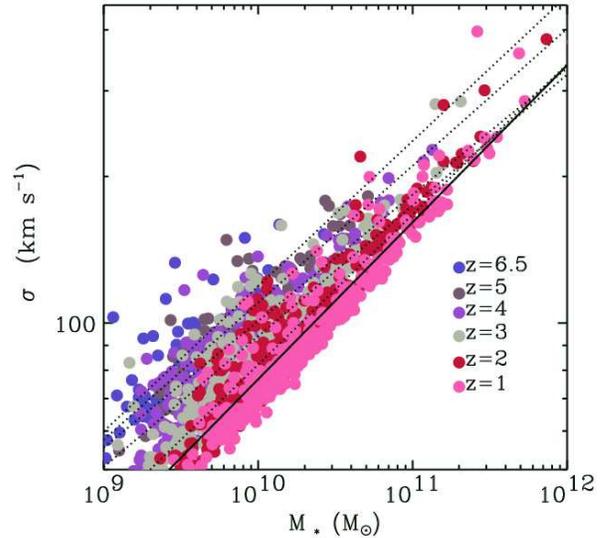}}
\caption{The stellar velocity dispersion $\sigma$ versus stellar
$M_{*}$ at $z=1$, $2$, $3$, $4$, $5$, $6.5$ (indicated by different
colors from blue to pink, the same ones as used in
Figs.~\ref{fig:D6acc_den}-\ref{fig:bhaf}). The best-fit power-law to
the trend is shown with a dotted line at each redshift, and with a
solid line at $z=1$. The dispersion $\sigma$ at fixed $M_{*}$
increases with increasing redshift, which can be interpreted as a weak
evolution in the Faber \& Jackson relation.
   \label{fig:sigma_mstar}}
\end{figure}

\begin{figure}
  \resizebox{9cm}{10cm}{\includegraphics{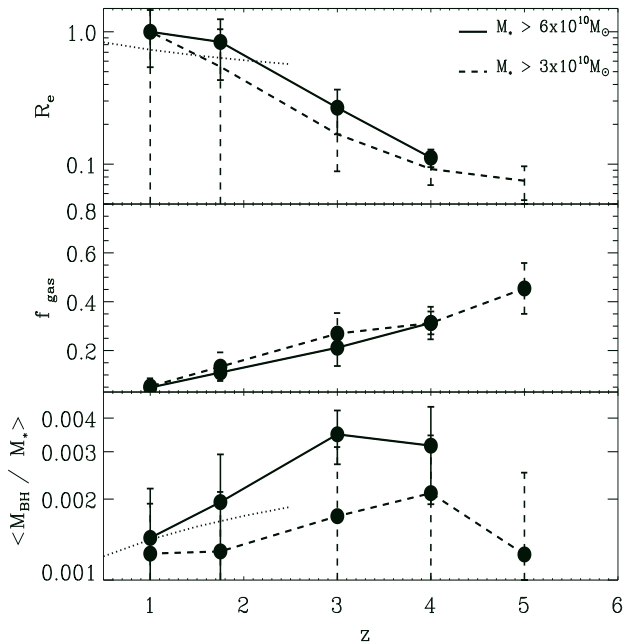}}
    \caption{The redshift evolution of the projected half mass radius
$R_{e}$ (top panel), cold gas fraction $f_{\rm gas}$ (middle panel)
and characteristic black hole to stellar mass ratio $\left <M_{\rm
BH}/M_{*}\right>$ (bottom panel), for systems with $M_{*} > 3\times
10^{10} \Msun$ (dashed lines) or $M_{*} > 6 \times 10^{10} \Msun$
(solid) in the {\em BHCosmo} run. These threshold values were chosen
to compare with the observed evolution determined by
\citet{Trujllo06}. The increase in cold gas content in high redshift
progenitor hosts, and the trend to more compact systems with an
increasing ratio of $M_{\rm BH}/M_{*}$ at a fixed stellar mass is
consistent with the recent results of \citet{Trujllo06} and
\citet{Peng2006}, as well as the BHFP \citep{Hopkins2007a}.
    \label{fig:fgasre}}
\end{figure}

\begin{figure*}
\begin{center}
\resizebox{9.5cm}{!}{\includegraphics{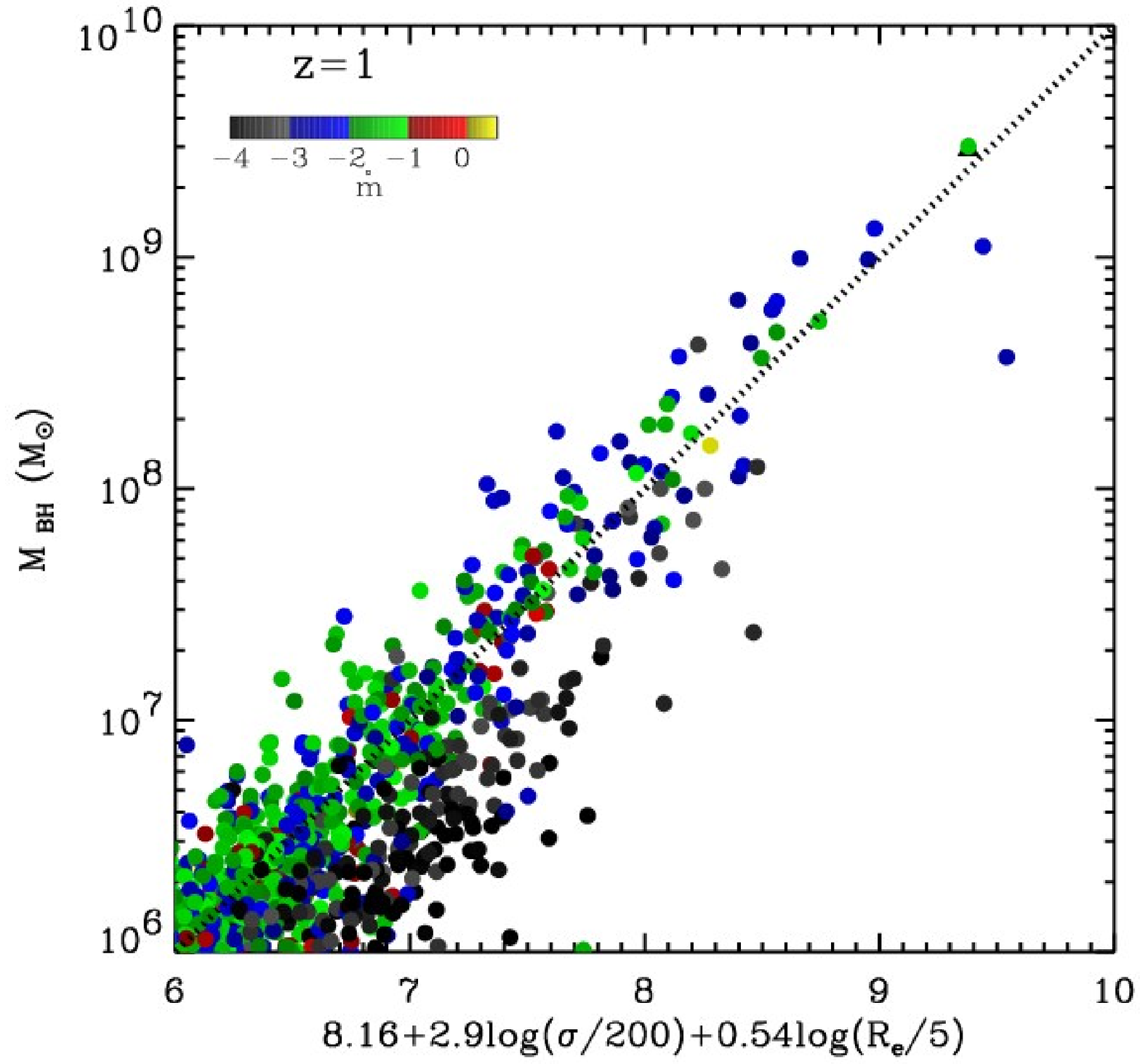}}%
\hspace*{-1cm}\resizebox{9.5cm}{!}{\includegraphics{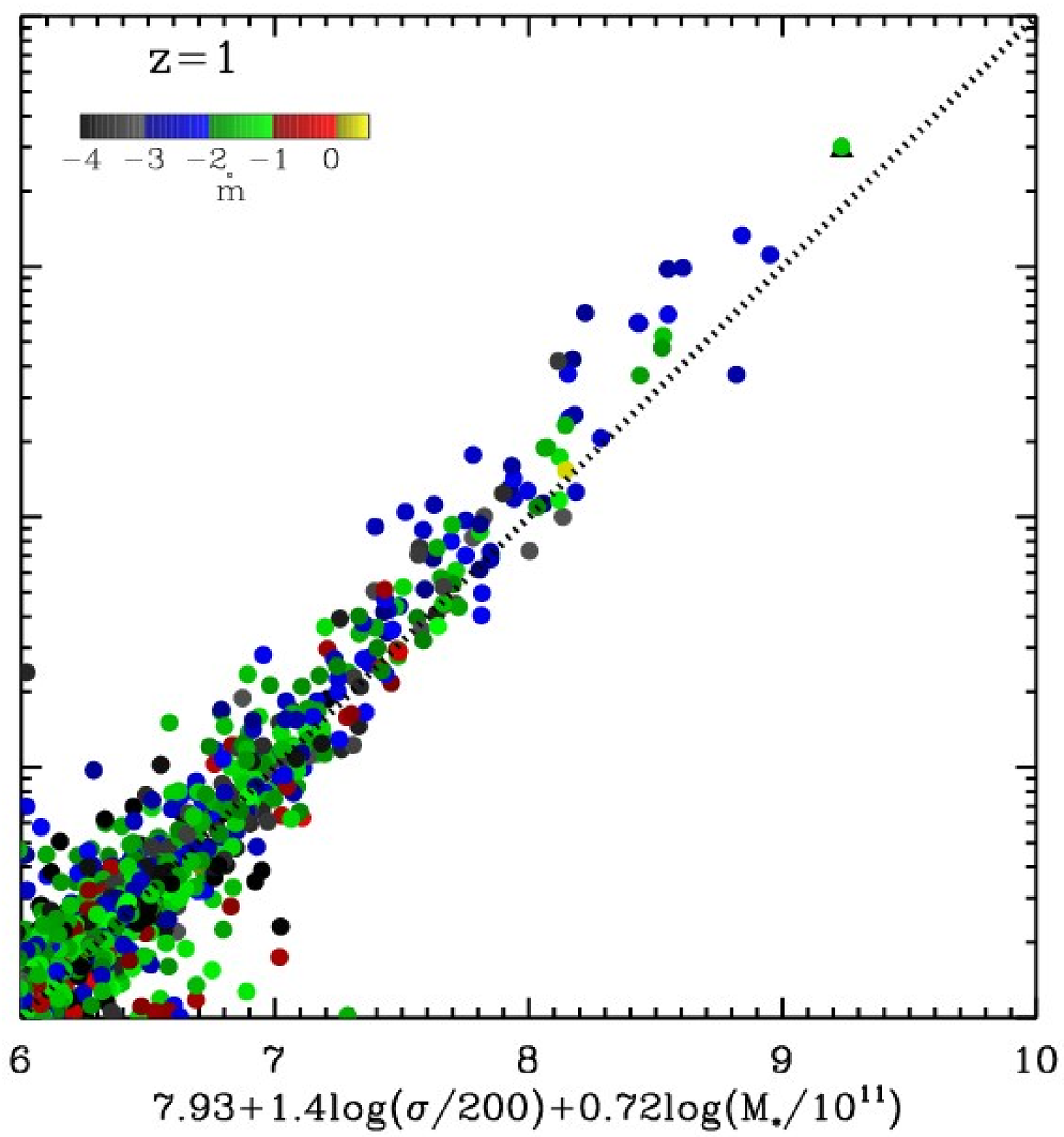}}
\end{center}
 \caption{Simulation prediction for two projections of the `Black Hole
Fundamental Plane' (BHFP) relation at $z=1$, in terms of $\sigma$ and
$M_{*}$ (left panel), and $\sigma$ and $R_e$ (right panel).  We
compare with the best-fit relations from \citet{Hopkins2007a}, shown
as dotted lines. The simulation agrees well with the conjecture of a
BHFP, which confirms the overall trends we have found in the $M_{\rm
BH}-\sigma$ and $M_{\rm BH}-M_{*}$ relations. This likely owes to
the scatter in the measurements of both $\sigma$ and in particular
$R_e$ which are noisy measurements in the simulations. The colors
indicate accretion rate values (as in previous figures and indicated 
in the color bar).}
    \label{fig:fplane}
\end{figure*}

The overall normalization of the $M_{\rm BH}-\sigma$ relation evolves
weakly with redshift when compared to the TO2 result for the present
epoch. Below $z \sim 3$, there is virtually no evolution (bearing in
mind the weak trend discussed above for $\sigma > 150 \kmps$), while
we find a weak drift of the normalization beyond $z\sim 3$.  A rough
parameterization of this evolution in the normalization of the $M_{\rm
BH}-\sigma$ relation is given by $ M_{\rm BH}/\sigma^4 \propto
(1+z)^{\alpha}$ with $\alpha =-0.2$.

These results for an overall modest evolution in the normalization are in
qualitative agreement with those from an analysis of isolated galaxy merger
simulations by \citet{Robertson2006a}.  We note that in the latter study a
redshift scaling of galaxy properties and a specific set of structural
properties of the merging galaxies had to be assumed, which introduced an
important systematic uncertainty. While having lower numerical resolution per
merger, the simulation we analyze here eliminates this caveat by providing a
fully self-consistent cosmological history for the formation and evolution of
galaxies and their black holes (albeit at a lower spatial resolution). This
provides an important confirmation of the analysis of \citet{Robertson2006a},
who however did not find evidence for an evolution of the slope at the high
mass end of the $M_{\rm BH}-\sigma$ relation.



\subsubsection{The evolution of $M_{\rm BH} -M_{*}$}
In Figure~\ref{fig:mbhmstar}, we show the $M_{\rm BH} -M_{*}$ relation
from the simulations at $z=1$, $2$, $3$, $4$, $5$ and $6.5$, alongside
the local observational relation determined by \citet[][thick grey
lines]{Haring2004}. Our best-fit relation at each redshift is shown by
a solid line while the dotted lines in each panel show results for the
other redshifts. Table~\ref{tab:mbhmstar} gives the slope $c$,
normalization $d$, and dispersion $\Delta_m$ for our best-fit
relations of the form
\beq \log \left(\frac{M_{\rm BH}}{\Msun}\right) = c \log
\left(\frac{M_{*}}{10^{11} \Msun}\right) + d.
\label{eqn:mbhmstarfit}
\eeq As before, our fitted values for $c$ and $d$ are intended to
indicate general trends in the evolution rather than to be used as
statistically rigorous measurements.  The observed relationship
\citep{Haring2004} has a slope $c=1.12$ and normalization $d=8.2$.

Overall there appears to be only limited evolution in the $M_{\rm BH} -M_{*}$
relation, but there is a slight steepening at $z=2-4$.  To highlight this
trend, we restrict our fits to the high mass end with $M_{*}\ge 5\times
10^{10}\Msun$ (dashed line in Fig.~\ref{fig:mbhmstar}). In this range, the
relation is significantly steeper, implying slopes $c_s \sim 1.9$ at $z=3-4$
and $\sim 1.5$ at $z\sim 2$.  This is more significant than the evolution
found in the slope of $M_{\rm BH}-\sigma$, and implies that there is some
evolution in the ratio of black hole mass to stellar mass relative to the
local observations. More precisely, systems with $M_{*} \simgt 10^{10} \Msun$
have larger black hole masses at fixed $M_{*}$ than at $z=1$, where the ratio
is in good agreement with the relation observed at the present epoch.  This
trend of an increasing $M_{\rm BH}/M_{*}$ ratio as a function of redshift
appears consistent with the recent measurements of high redshift (up to $z
\sim 3.5$) BH masses and host luminosities by \citet{Peng2006},
as well as the BHFP \citep{Hopkins2007a}. We will
further analyze this effect in \S~\ref{sec:fgas}.

\begin{table}
\begin{center}
\caption{Parameters of best-fit $M_{\rm BH} -M_{*}$ relations}
\label{tab:mbhmstar}
\begin{tabular}{ccccc}
\hline\hline\\
$z$  &  slope $c$ & normalization $d$ & scatter $\Delta$ & c$_{s}$\\
 \hline\\
1 ....... &  $1.18\pm0.02$ & $8.10\pm0.03$ & 0.03 &1.2 \\
2 ....... &  $1.23\pm0.03$ & $8.09\pm0.03$ & 0.04 & 1.5 \\
3 ....... &  $1.25\pm 0.04$ & $8.04\pm0.04$ & 0.06 & 1.9 \\
4 ....... &  $1.30\pm 0.05$ & $8.04\pm0.05$ & 0.07 &1.9 \\
5 ....... &  $1.17\pm 0.10$ & $7.90\pm0.10$ & 0.14 & 2.0 \\
6.5 ..... &  $1.01\pm 0.22$ & $7.78\pm0.25$ & 0.34 & -- \\
\hline\\
\end{tabular}
\end{center}
\vspace{-1cm}
\end{table}

\subsection{Evolution of $\sigma$, gas fraction, and $R_e$,
 at fixed stellar host mass}
\label{sec:fgas}
We now analyze some of the physical properties of the host galaxies
and their evolution with redshift to investigate the physical origin
for the trends we have found in the $M_{\rm BH}-\sigma$ and $M_{\rm
BH} -M_{*}$ relations.  Figure~\ref{fig:sigma_mstar} shows the stellar
velocity dispersion $\sigma$ versus the stellar mass $M_{*}$ for each
galaxy as a function of redshift. The dotted lines and the solid line
(for $z=1$) show our best-fit relations. At a fixed $M_{*}$, the
velocity dispersion $\sigma$ increases with increasing redshift. This
is consistent with the results from the merger remnants in
\citet{Robertson2006a} and reflects changes in the structural
properties of stellar spheroids, which become smaller towards higher
redshift.

In Figure~\ref{fig:fgasre}, we show the projected half mass radius $R_e$, the
cold gas fraction $f_{\rm gas}$ (a proxy for the disk gas fraction), and the
$\left<M_{\rm BH}/M_{*}\right>$ ratio as a function of redshift. We show
results for systems with $M_{*} \ge 3 \times 10^{10} \Msun$ (dashed lines) and
$M_{*} \ge 6 \times 10^{10} \Msun$ (solid lines). We find higher cold gas
fractions in higher redshift systems with $f_{gas} \propto
(1+z)^{1.5}$. The increased gas content, dissipation rate in high redshift
progenitor is consistent with them being more centrally concentrated, as
shown by the trend of decreasing $R_{e}$ and increasing $\sigma$ with
increasing redshift, leading to larger ratios $M_{\rm BH} /M_{*}$ at high
redshifts, with $M_{\rm BH} /M_{*} \sim (1+z)^{0.5}$, at least over these
ranges of $M_{*}$. Larger values of $\sigma$ at fixed $M_{*}$ at high redshift
are also consistent with an inverse overall trend (e.g.; $(1+z)^{-0.2}$) in
the evolution of the $M_{\rm BH}-\sigma$ relation.

The above results are consistent with the general picture for a merger-driven
quasar growth we have developed based on our galaxy merger
simulations~\citep{DiMatteo2005, Springel2005a, Robertson2006a, Hopkins2005,
Hopkins2006a, Hopkins2007a} , which also implies that the $M_{\rm BH}-\sigma$
and $M_{\rm BH}-M_{*}$ relations are connected to each other. This connection
is most clearly demonstrated by the `black hole fundamental plane' (BHFP)
relation discussed by \citet{Hopkins2007a}, which arises from the joint
formation process of spheroids and massive black holes in mergers. Our
cosmological simulations also should agree with the BHFP if most of the BH
growth is associated with mergers.  In Figure~\ref{fig:fplane}, we show our
simulation measurements for the BHFP and compare it to the the best-fit from
\citet{Hopkins2007a}, shown as a solid line.  We find very good agreement with
the predictions for the BHFP by \citet{Hopkins2007a}, in both of its
formulations, i.e.~for $R_e$ and $\sigma$ or $M_{*}$ and $\sigma$, although
the former shows a larger scatter than the latter, owing to the noisy
measurements of $R_e$ in the cosmological simulation.

\section{The first and the most massive black holes}
\begin{figure*}
\begin{center}
\resizebox{17.0cm}{!}{\includegraphics{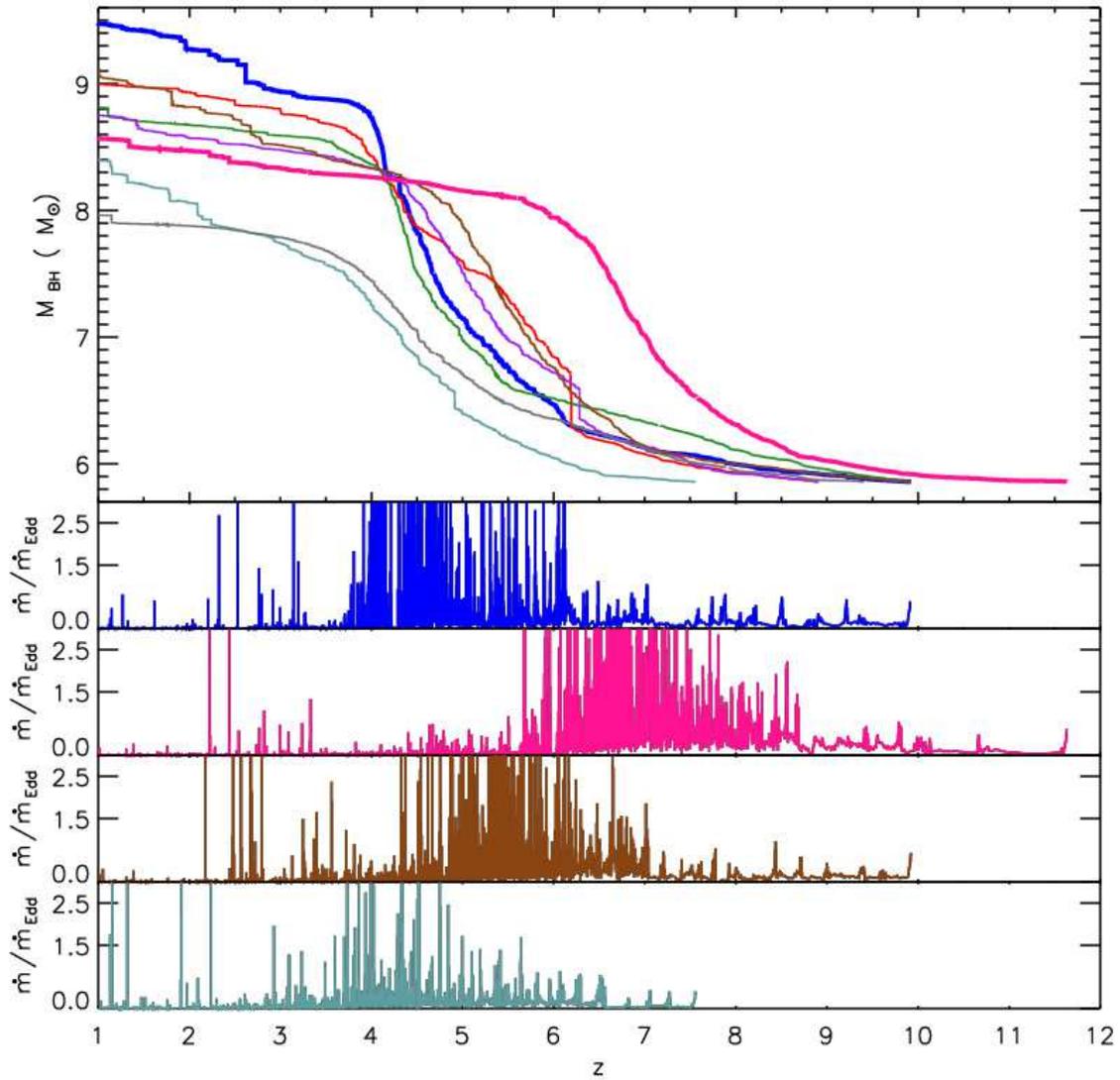}}\vspace*{-1.8cm}
\end{center}
 \caption{Individual mass assembly and accretion rate histories for
the six most massive black holes and two intermediate mass BHs (chosen
randomly) in the {\it BHCosmo} run. The top panel shows the black hole
mass as a function of redshift while the bottom four panels give the
detailed accretion rate history for the black holes with the
corresponding colors in the top panel. The first supermassive black
hole forming in the {\it BHCosmo} run, and the most massive one at the
end of the run, are shown with thicker lines in pink and blue,
respectively. From the bottom panels we see that phases of high
Eddington accretion occur at different times of the history of
different black holes.
\label{fig:accrhistD6}}
\end{figure*}
In the previous sections we have discussed the predictions from our
simulation for the global history of black hole mass assembly in
galaxies from the high redshift Universe to today. We have compared
the predictions for the evolution of the black hole mass and accretion
rate density to the history of the star formation rate density and
discussed the growth of the black hole mass function.  We have also
discussed the $M_{\rm BH} - \sigma$ and $M_{\rm BH} -M_{*}$
relationships as a function of time, and examined which physical
properties drive their cosmological evolution.

However, our simulation methodology not only allows us to make statistical
statements about the black hole population. Rather, it can also be used to
study the detailed growth history of individual black holes, from the moment
they are seeded to today, which provides a particularly powerful way to follow
the evolution of black holes over cosmic time. The gas which fuels black hole
activity ultimately has its origins in the intergalactic medium, draining
along filaments into forming galaxies. Because of this, BH radiative histories
are directly linked to the formation of large-scale structure in the Universe,
from supercluster scales down to the immediate environment of host galaxies.
Being able to follow these large-scale processes and their impact on
individual black holes self-consistently is the key to a qualitatively better
level of understanding. For example, we have the tools to examine what turns a
small black hole into a supermassive one at $z=6$, or whether some BHs grow
hardly at all after this initial phase (as we shall see below in some
examples). We can ask how quasar lifetimes are related to their clustering,
how important BH mergers versus gas accretion are in the cosmological growth
of BH mass, what BH light curves look like in detail over a Hubble time,
and how specific outbursts correlate with galaxy and cluster merger and
accretion events.

In future work we will return to address such questions in more detail; for
example in \citet{Colberg07} we present an extensive study of BH environments
as the universe evolves.  For now we will show that there are some striking
inferences to be drawn from following the detailed histories of even a few
black holes.  As an example we choose to focus on the formation and fate of
the first large black holes that form in our simulation, as well as the
properties of their hosts and their descendants. This allows us to identify
the prerequisite conditions for the growth of supermassive black holes already
in the early universe.

\begin{figure*}
\begin{center}
\resizebox{17.0cm}{!}{\includegraphics[angle=270]{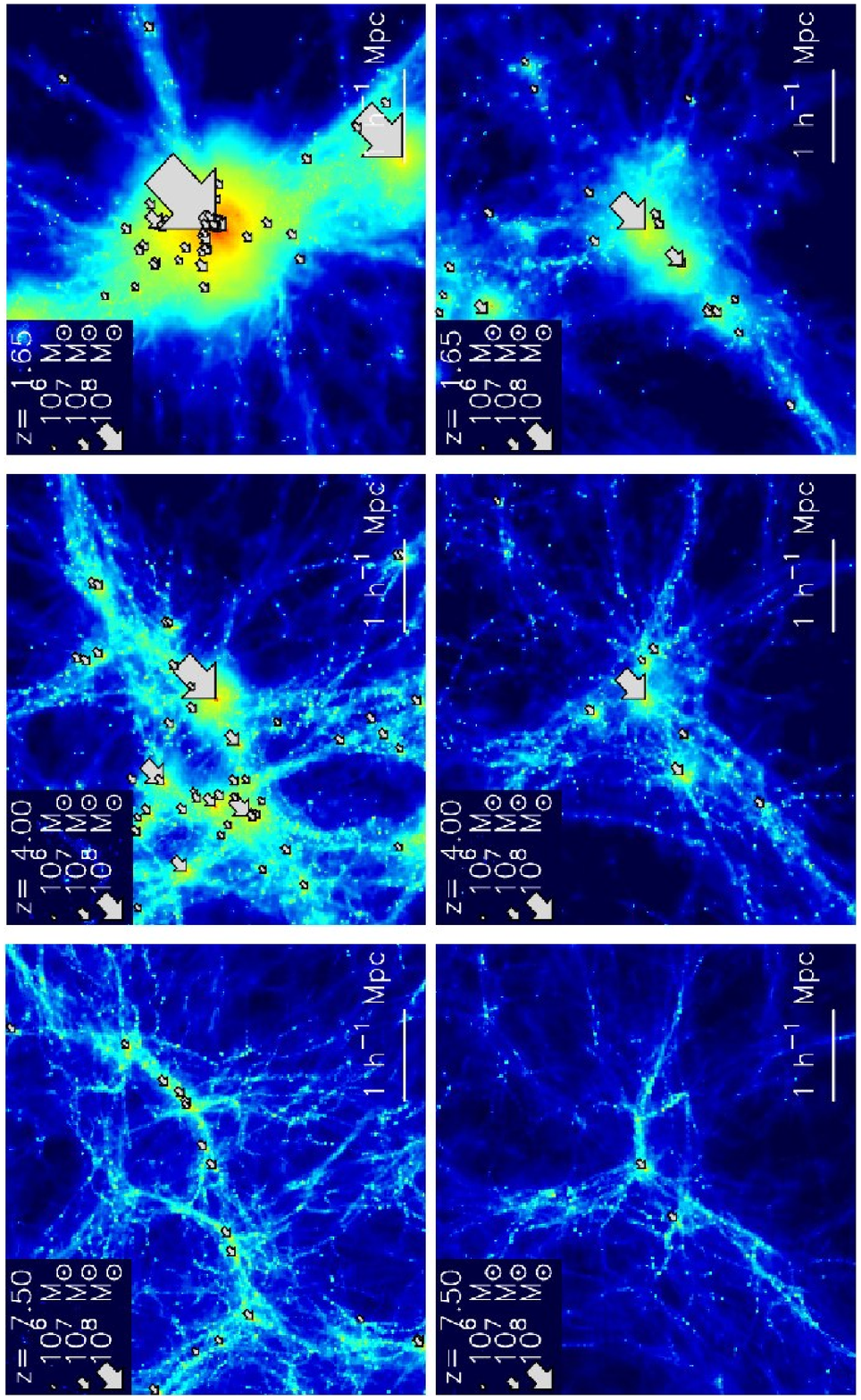}}
\end{center}
 \caption{Time evolution of the environment around the host galaxy of
the first supermassive massive black hole at $z=6$ (bottom panels),
and of the most massive BH at $z=1$ (top panels) in the {\it BHcosmo}
run. The location and masses of the supermassive BHs are marked by
arrows of different size, as labeled. While the bottom system hosts
the most massive black hole at high redshift, it does not end up
hosting also the most massive black hole at the center of the largest
galaxy at low redshift. Instead, the system shown in the top panels
overtakes it in growth at intermediate redshifts, when it is formed in
the highest density region in the simulation, which is a protocluster
region.
    \label{fig:2bigbhs}}
\end{figure*}

Observations of luminous SDSS quasars at redshifts as high as $z \sim
6$ \citep{Fan2003} present a number of challenges for models of high
redshift quasar and galaxy formation. Their low space density suggests
that they reside in the rarest dark matter density peaks at this early
epoch, yet the apparent lack of companion galaxies in the field has
been used to argue that these quasars reside in far more common halos
\citep{Carilli2004, Willott2005}. Extensive follow-up observations of
the highest redshift ($z=6.42$) quasar, SDSS J11148+5251, are starting
to constrain the properties of its host galaxy \citep{Walter2004}.  CO
observations indicate a relatively small stellar spheroid, however,
the existence of CO, iron and carbon emission indicates a heavily
enriched ISM and vigorous star formation \citep{Maiolino2005}. In any
case, rapid growth is clearly required to produce large supermassive
black holes and spheroids in the $\sim 800$ million years available to
$z\sim 6$.  Unsurprisingly, the type of object associated with the
'first quasar' is therefore still a matter of debate.


The relatively small volume of our simulation prevents us from directly
addressing the problem of the `first quasars', simply because their space
density is so low. Properly identifying one of the rare host systems requires
at the very least box-sizes of $500\,h^{-1}{\rm Mpc}$
\citep{SpringelMillennium2005}, or more adequately $1\,h^{-1}{\rm Gpc}$
\citep{Li2007}.  Nevertheless, our simulation allows us to examine the nature
of the environments and the hosts where exponential growth of BHs can first
occur at early times. This has some relevance to the problem of where the
first quasars are likely to form.

In order to retrieve the required information for all the black holes in the
simulation we have constructed full merger trees for each of them. This allows
us to track the growth and accretion history of individual black holes of
different masses, make detailed lightcurves from their accretion histories,
and study the properties of their host galaxies.

\subsection{Individual BH mass and accretion rate as function of $z$}
In Figure~\ref{fig:accrhistD6}, we show the accretion histories of a
number of black holes in the {\it BHCosmo} run (top panel) and some
examples of their corresponding accretion rate histories in the bottom
four panels. The sample in the top panel consists of the six most
massive black holes in the simulation and two randomly chosen ones
with an intermediate mass at $z=1$. The four bottom panels show the
accretion rate in units of Eddington for four black holes with
corresponding line colors, including the most massive black hole at
$z=6$, the most massive and second most massive at $z=1$, and finally
one of the intermediate mass ones.  Note that this is a small
sub-sample of the several thousand black holes in the simulation. We
focus here on the evolution of the most massive ones, and in
particular on their early formation and the fate of their descendants.
In \citet{Colberg07}, we study a complete sample of individual BH
histories as a function of environments for the entire simulation.

The first interesting result shown in Figure~\ref{fig:accrhistD6} is
that even in this small volume conditions exist that are conducive for
exponential black hole growth, as required by the presence of quasars
at $z\sim 6$ with the large observed masses. The evolution of the mass
of the largest black hole in our simulation at early times is shown by
the thick pink line.  The corresponding accretion rate history (in the
same color) shows a rapid succession of numerous phases of high
accretion at the critical Eddington rate, between $ 5 \simlt z \simlt
7.5$. Below $z\sim5$, the black hole is drastically more quiescent,
except for a couple of sporadic Eddington phases at around $z\sim 2$.
Interestingly, this first most massive black hole at $z \sim 6$ is not
the most massive one at $z=1$. Instead, it is `overtaken' in growth by
another black hole shown with the blue thick line, which is the most
massive black hole at the end of the simulation.  The Eddington growth
phases for this objects start at $z < 6$ and extend all to way to
$z\sim 3.5$. By $z \sim 4.5$, the black hole mass in this system
catches up with that of the first massive black hole, and then keeps
growing exponentially to $z\sim4$.

In Figure~\ref{fig:2bigbhs}, we show projected gas density maps of the
host galaxies and the surrounding structures (at redshifts $z=7.5$,
$4$, and $1.6$) for two of these black holes, the one that is most
massive at $z=1$, and the one that is most massive at $z=6$.  This
provides good clues for the origin of the evolutionary difference
between these two systems. Although the hosts of both black holes at
$z=7.5$ are halos of similar mass, $\sim 10^{10} \Msun$, one of them
lies at the intersection of three small filaments, leading to
efficient gas cooling and a high star formation of $\sim 100
\Msun$yr$^{-1}$. The host galaxy grows rapidly until $z=6$ but then
its gas supply dwindles, and at $z\sim4$ it is overtaken by the other
black hole.  This latter one lies on a very massive filament, which
grows even more vigorously at this later time than its host galaxy
(which has a SFR of $\sim 1000 \Msun$yr$^{-1}$ at $z\sim 2$). In this
way a large stellar spheroid is produced around the BH, which
eventually will end up at the center of a rich cluster of galaxies.
Our results therefore show that a massive black hole that is found in
the cD of large galaxy cluster at late times was not necessarily the
most massive one at $z=6$, as it has been often assumed in the
literature \citep[e.g.][]{SpringelMillennium2005}. Since the growth
history of black holes is intertwined with the non-linear processes of
structure formation, individual growth histories of BHs can be complex
and need not preserve the rank order in a group of BHs that start out
with similar masses.

\begin{figure}
\hspace{-0.5cm}
\resizebox{8.5cm}{!}{\includegraphics{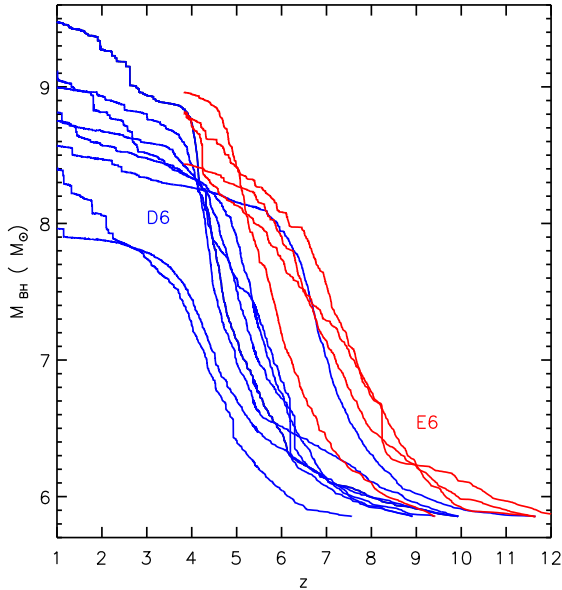}}
 \caption{Comparison of individual black hole mass histories in the
{\it BHCosmo} run (blue lines, D6) and in the larger simulation volume
of the E6 run (red lines). The growth of the first supermassive black
holes at $z\sim 6-7$ is more widespread in the larger volume of the
E6. The 'catch-up' of larger black holes that form later and in higher
density regions (see text) can also be seen, and is comparable to the
case shown in Fig.~\ref{fig:accrhistD6}.
    \label{fig:accrhistD6E6}}
\end{figure}

\begin{figure*}
\begin{center}
\resizebox{17.0cm}{!}{\includegraphics{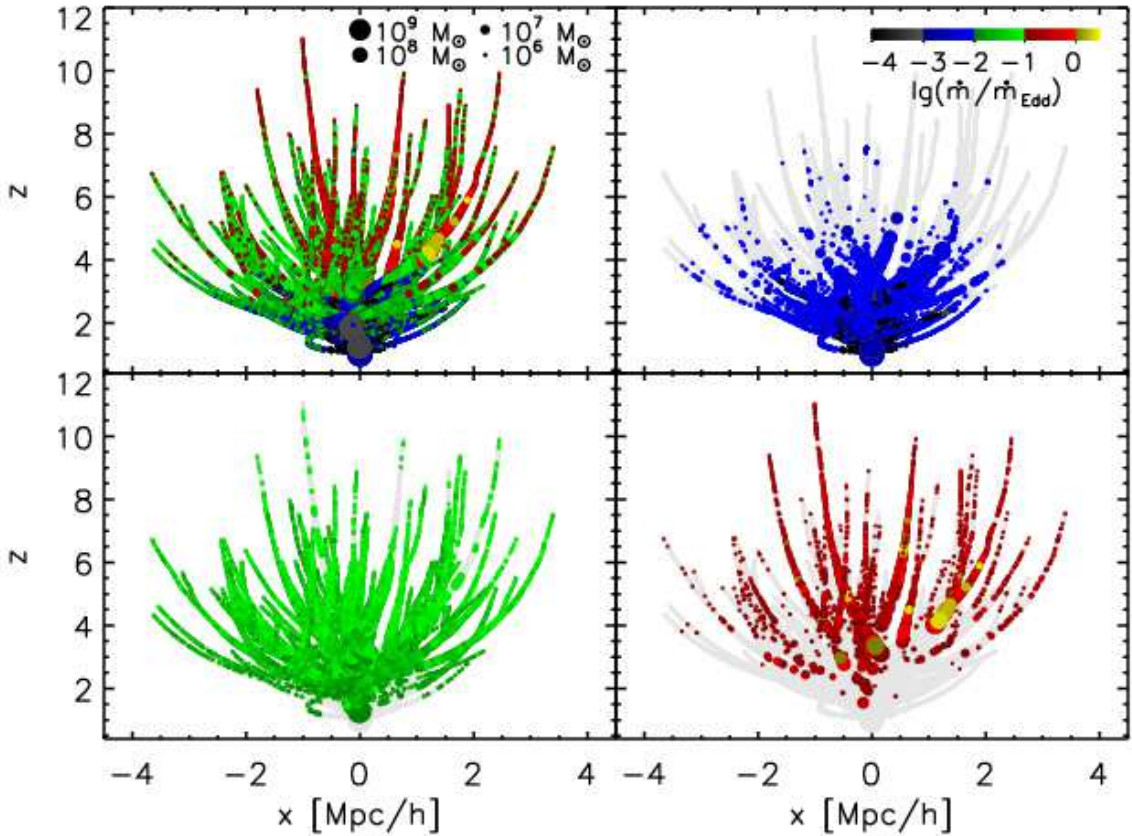}}%
\vspace*{-0.8cm}
\end{center}
 \caption{Black hole merger history trees shown as a function of
redshift ($y$-axis) and position (along the $x$-axis), for the most
massive black hole at $z=1$ in the {\it BHCosmo} run.  The full tree
is shown in the top left panel, while the other three panels split up
the tree according to accretion rate in units of Eddington, as
indicated by the colors. The black hole mass is given by the size of
the symbols, as labeled.
\label{fig:tree1}}
\end{figure*}

\begin{figure*}
\begin{center}
\resizebox{17.0cm}{!}{\includegraphics{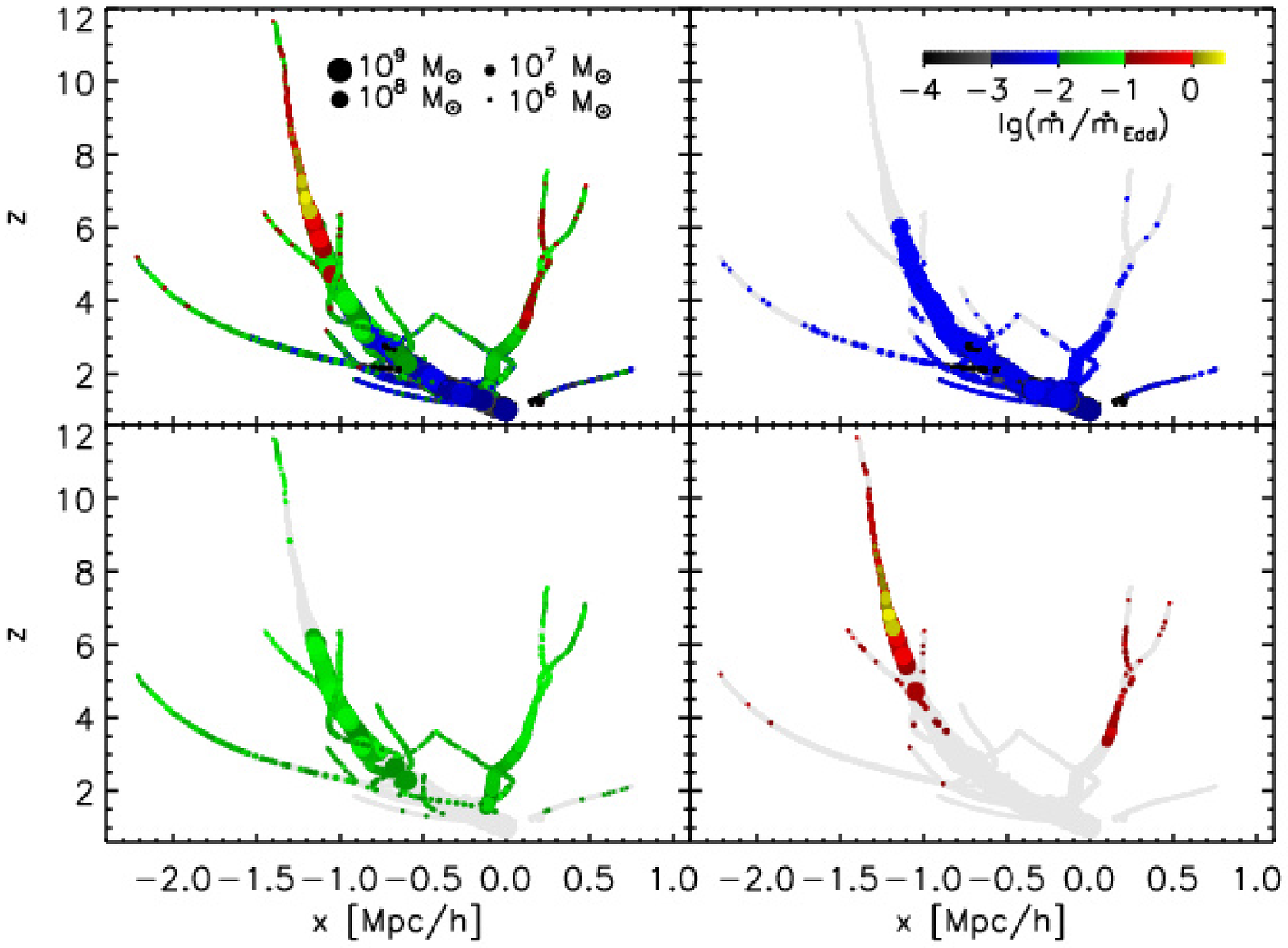}}
\vspace*{-0.8cm}
\end{center} 
\caption{Black hole merger history tree shown as a function of redshift
($y$-axis) and position ($x$-axis) for the first high-redshift massive black
hole that forms in the simulation. The full tree is shown in the top left
panel, while the other three panels split up the tree according to accretion
rate in units of Eddington, as indicated by the colors. The black hole mass is
given by the size of the symbols, as labeled. The evolution of this system is
qualitatively very different from the one shown in Fig.~\ref{fig:tree1} Note
that discreteness in the outputs for black holes that remain inactive cause
the gap in the rightmost branch of this tree.
    \label{fig:tree6}}
\end{figure*}

Further confirmation of this result from the {\it BHCosmo} simulation
come from our E6 run (see Table~\ref{tab:simul}), which includes the
same physics and model parameters as the {\it BHCosmo} simulation but
samples rare halos better, thanks to its larger box-size of $50
\,h^{-1}{\rm Mpc}$ on a side.  In this larger volume, there are a
handful of more examples of early exponential black hole growth
between $8\simlt z \simlt 6$, as shown by the red black hole mass
growth curves in Figure~\ref{fig:accrhistD6E6}.  Consistent with our
previous finding, there are again black holes that 'catch-up' in
growth to the most massive black hole at some earlier time. We note
that we have run this larger volume simulation with the specific
purpose of checking the {\it BHCosmo} results for the first
supermassive black holes by obtaining better statistics for large
halos at $z\simgt 6$. For this reason, this simulations was not
evolved beyond $z=4$.

\subsection{BH merger trees for the first and the most massive black hole}
Figures~\ref{fig:tree1} and \ref{fig:tree6} show two example black
hole merger trees, one for the most massive black hole at $z=1$, and
one for the $z=1$ descendant of the first supermassive black hole at $z
\sim 6$. All progenitor black holes that merge together to build up
the final BH are included in the trees.  The black hole masses along
the tree are represented by different sizes of the circles, while the
color encodes their accretion rate in units of the Eddington rate.  In
both figures, the top-left panel shows the full BH merger tree,
whereas the top-right and bottom two panels split the tree according
to accretion rate of the BHs.

Inspection of these two BH trees immediately reveals a complex and
rich merger history for the most massive black holes at $z=1$
(Fig.~\ref{fig:tree1}), as opposed to the comparatively isolated
evolution of the most massive object at $z=6$, which only features
three major branches for its tree (Fig.~\ref{fig:tree6}). The large
majority of the mass in the latter case is built up by early phases of
critical gas accretion at $z \simlt 6$. On the other hand, in
Figure~\ref{fig:tree1}, Eddington accretion phases are spread out
along many branches of the tree and cover a much larger range of
redshifts.  At redshifts $z < 3$, the progenitors are preferentially
accreting at low Eddington rates, so that the residual growth of the final
black hole mass increasingly occurs from ``dry'' mergers with other
black holes, consistent with our earlier results for the global black
hole growth.

\section{Summary and discussion\label{sec:conclusions}}
In this paper we have investigated the coupled formation and evolution
of black holes and galaxies using state-of-the-art cosmological
hydrodynamic simulations of the $\Lambda$CDM model. For the first
time, we have incorporated black hole growth and associated feedback
from quasar activity self-consistently in cosmological hydrodynamic
simulations. Our approach has been based on the methodology recently
developed and tested in simulations of galaxy mergers
\citep{DiMatteo2005, Springel2005a}. In this initial paper we have
focused on investigating the model predictions for (i) the global
history of black hole mass assembly and its relation to the history of
star formation, from high redshift to the present, for (ii) the
evolution of the BH mass function and accretion rate function and its
connection to the observational ``downsizing'' phenomenon, for (iii)
the correlations between black hole mass, velocity dispersion, and
stellar mass of host galaxies, and finally, for (iv) the formation and
fate of the first quasars and the properties of their hosts.

An important and highly encouraging first result has has been that the
cosmological black hole mass density $\rho_{\rm BH}$ predicted by our
high-resolution simulation reproduces the locally measured value and
its extrapolation to higher redshift ($z< 2.5$), inferred from
integrating the X-ray luminosity functions \citep{YuTremaine2002, Marconi2004,
Shankar2004}, or, more directly, the bolometric quasar
luminosity function \citep{Hopkins2007b}.
We predict a steep evolution of $\rho_{\rm BH}$ as a
function of redshift, with a rise that is more rapid than that of the
star formation rate. At $z > 3$, $\rho_{\rm BH}/\rho_{*} \propto
(1+z)^{-3}$ whereas there is at most weak evolution in the ratio
$\rho_{\rm BH}/\rho_{*}$ at $z \lesssim 3$. Similarly, whilst the star
formation rate density $\dot\rho_{*}$ broadly tracks the BHAR density
$\dot\rho_{\rm BH}$ below $z\sim 2.5-3$, their ratio evolves steeply
at higher redshifts, as $\dot\rho_{\rm BH}/ \dot\rho_{*} \propto
(1+z)^{-4}$.  The SFR density peaks earlier than the BHAR density and
exhibits a more gradual evolution with redshift compared to the BHAR.
Only at redshifts below the peak of the BHAR, the BHAR and SFR rate
densities start tracking each other.

Our results for the evolution of the BHAR density are broadly
consistent with constraints obtained by \citet{Hopkins2007b} from a
comprehensive analysis of a large sample of observational data sets,
which allowed them to synthesize the evolution of the bolometric
quasar luminosity density with redshift and to show that the
luminosity density indeed peaks at $z\sim 2-3$, with a sharp drop
towards higher redshifts.

We have shown that the growth of the black hole mass function shows
signs of `anti-hierarchical' behavior, or `downsizing'.  The high mass
end of the BH mass function is established at comparatively high
redshift and then significantly slows down in evolution below $z \sim
2$, where only the abundance of intermediate mass BHs is still growing
appreciably.  We have found more direct evidence for black hole
``downsizing'' by inspecting the distribution of accretion rates, in
unis of the critical Eddington rate. The accretion rate function
shifts from a narrow distribution dominated by high Eddington rates at
high redshift to a broad distribution with a small fraction of
Eddington accretion for low redshifts ($z < 3$). High mass black holes,
forming in the high density peaks at high redshift are built up
rapidly by vigorous accretion. However, the effects of gas depletion
and AGN feedback drive a strong decline in the accretion rate of
massive black holes towards late times, while during these later times
the peak of the accretion activity shifts to progressively smaller
mass scales.


The BH masses of our simulated galaxies are strongly correlated with the
stellar velocity dispersions and stellar masses of their host galaxies, and
the correlations agree remarkably well with the local $M_{\rm BH} - \sigma$
and $M_{\rm BH}-M_{*}$ relationships, over a large dynamic range. We
previously showed with simulations of isolated galaxy mergers
\citep{DiMatteo2005, Robertson2006a, Hopkins2007a} that our AGN feedback
prescription leads to a self-regulated BH growth that can explain the 
$M_{\rm BH} - \sigma$ relation. 
It is highly reassuring that the same simulation
model, with an unmodified feedback efficiency $\epsilon_f$ reproduces the
observed $M_{\rm BH} - \sigma$ in full cosmological simulations as well. We
emphasize that the free parameter $\epsilon_f$ sets the normalization of the
obtained relationship, but the slope and scatter of the relation obtained from
the simulations are not adjustable and a non-trivial consequence of the
self-regulated BH growth.

Our simulation also suggests a weak evolution with redshift of the
normalization of the $M_{\rm BH}-\sigma$ relation, as $M_{\rm BH}
/\sigma \sim (1+z)^{-0.2}$. However, we find that this evolutionary
trend is sensitive to range of masses being probed. When we focus on
the better resolved more massive systems, there appears to be some
mild evolution in the slope of the BH scaling relations at high
$\sigma$ and for high $M_{*}$. In particular, we find a trend of
increasing $M_{*}/ M_{\rm BH}$ with redshift, in agreement with recent
direct estimates of the BH to host stellar mass ratio at high redshift
by \citet{Peng2006}. This trend is accompanied by an increase in the
cold gas fraction in the host galaxy, and a smaller $R{_e}$ at a fixed
stellar mass.  These results are consistent with the suggestion that
the $M_{\rm BH} - \sigma$ and $M_{\rm BH}-M_{*}$ relations are
projections of a more basic correlation,
the black hole fundamental plane \citep{Hopkins2007a}, 
$M_{\rm BH} \propto \sigma^3
R_{e}^{0.5}$, and the interpretation that gas-rich systems at high
redshift produce more dissipative mergers that lead to more
concentrated BH hosts.

Our findings for the BH scaling relations appear fully consistent with a
scenario where quasar activity is driven by galaxy mergers, as suggested by
our simulations of isolated mergers \citep{DiMatteo2005, Springel2005a,
Robertson2006a, Hopkins2006a, Hopkins2007a}. While gas is available for star
formation over a large range of mass scales, it can only gets to central
regions of galaxies in large amounts as a result of the inflows that accompany
major mergers. At the same time, the mergers produce spheroids, which together
with the growth-limiting quasar feedback establishes the $M_{\rm BH} - \sigma$
relationship.  The detailed time evolution of the quasar activity in an
individual galaxy depends on the angular momentum of the gas and the disk size
and morphology. It is clear that the limited spatial resolution of our
cosmological simulation is a severe limitation on how faithfully this can be
tracked in our calculation, but because the final BH masses in merger remnants
are robustly predicted by our simulation methodology even at coarse resolution
this does not seriously affect the quantities we study here.

Our simulated volumes are too small to directly check if a population
of supermassive black holes as massive the $z\sim 6$ Sloan quasars can
form at high redshift. However, we do find that our simulation model
produces black holes at early times that spend a large fraction of
their time at high accretion rates close to their Eddington rates. We
therefore predict the presence of massive black holes at high
redshift. Using high-resolution simulations of multiple mergers that
were constructed to represent a high redshift merging tree of one of
the most massive protocluster regions in a $(1\,h^{-1}{\rm Gpc})^3$
volume, \citet{Li2007} have recently made the match with the Sloan
quasar much more explicit. They were able to show that the gas-rich
mergers expected in this rare overdensity can indeed grow a BH of mass
$\sim 10^{9}\,{\rm M}_\odot$ early enough.

We have also found that the first supermassive black hole in our simulation
forms in relative isolation, as a result of strong gas inflows and merging at
the intersection of large-scale filaments. However, by following this system
to lower redshift it turned out that it did not evolve into the most massive
black hole today. Instead, it was overtaken in growth by a black hole that
acquired most of its mass in major mergers in a high density region at lower
redshift. Such a change in the relative rank of BHs as a function of time
appears quite generic, as we explicitly checked with a simulation of larger
volume. This clearly complicates attempts to directly link high redshift
progenitor systems to low redshift descendants. As we have demonstrated, the
cosmological simulation methodology we introduced here provides however an
excellent tool for studying the evolution of the cosmic BH evolution. In the
present study, we focused only on the radiatively efficient accretion mode of
AGN, which is associated with quasar activity. For following AGN activity also
in clusters of galaxies, and hence down to $z=0$ in large volumes, we also
need to account radio activity, which becomes important in very massive halos
at low redshift. In \citet{Sijacki2007} we present an extension of our
simulation methodology that accounts for this physics, and we discuss results
obtained with this unified model for AGN feedback both for clusters of
galaxies and cosmological boxes.
 
Our cosmological approach to black hole formation enables us to follow the
fuel for black hole activity from its ultimate source, the early intergalactic
medium, and so link the large-scale structure and environments of black holes
directly with their growth. As a result, countless different avenues of
research are opened up and we plan to explore many in future work. For
example, in \citet{Colberg07} we use merger trees constructed for every black
hole in the simulation to carry out a systematic study of their environments
and histories and to determine the amount of black hole growth owing to
mergers versus gas accretion; the latter should be testable with upcoming
gravitational wave experiments. Future issues which will be addressed include
the nature of AGN clustering and its evolution and its relation to the
underlying dark matter clustering, predictions for quasar lifetimes and
luminosity functions, and how feedback from AGN may manifest itself on
large-scales and can be detected through the Sunyaev-Zeldovich effect. With
our new approach we can address the question of how the first black hole
formed and grew we will be able to make predictions for the ionizing
background owing to the first miniquasars, an important but currently uncertain
ingredient in current models of reionization.

\acknowledgements
We thank Rupert Croft for many discussions and reading the manuscript.
The simulations were performed at Carnegie Mellon University and the
Pittsburgh Supercomputer Center (PSC). 
This work has been supported in part through NSF AST-0607819.


\bibliographystyle{apj}
\bibliography{ms}

\end{document}